\RequirePackage{silence}
\documentclass[aps,prx,reprint,superscriptaddress,nofootinbib,10pt,longbibliography,floatfix]{revtex4-2}
\usepackage[T1]{fontenc}

\usepackage{parskip}
\usepackage{microtype}

\usepackage[colorlinks=true]{hyperref}
\hypersetup{
  colorlinks   = true, 
  urlcolor     = green!80!black, 
  linkcolor    = blue, 
  citecolor    = red!80!black 
}
\usepackage{orcidlink}

\makeatletter
\renewcommand{\bibinfo}[2]{%
  \ifstrequal{#1}{title}{\textit{#2}}{#2}}
\makeatother

\usepackage{amsmath,amssymb,amsthm,mathtools,bm,mathpazo}
\usepackage{algorithm,algpseudocode}
\floatstyle{plain}
\restylefloat{algorithm}
\floatname{algorithm}{ALGORITHM}
\makeatletter
\def\floatc@plain#1#2{\@makecaption{#1}{#2}}
\makeatother
\usepackage{graphicx,booktabs}
\usepackage[dvipsnames]{xcolor} 
\usepackage[nameinlink]{cleveref}
\usepackage{xparse}
\usepackage[normalem]{ulem}

\newcommand{\parTitle}[1]{\textit{#1 ---}}

\makeatletter
\newcommand{\columntoc}{%
  \begingroup
    \setlength{\parskip}{0pt}%
    \small
    \def\l@f@section{\addvspace{0.3em}}%
    \def\l@f@subsection{}%
    \let\columntoc@numberline\numberline@@sections
    \def\numberline@@sections##1{%
      \if\relax\detokenize{##1}\relax
        \leavevmode\ignorespaces
      \else
        \columntoc@numberline{##1}%
      \fi}%
    \par\noindent\rule{\columnwidth}{0.4pt}\par\smallskip
    {\centering\bfseries Contents\par}\smallskip
    \let\appendix\appendix@toc\@starttoc{toc}%
    \par\smallskip\noindent\rule{\columnwidth}{0.4pt}\par
  \endgroup
  \bigskip
}
\makeatother

\newtheorem{theorem}{Theorem}

\newtheorem{proposition}{Proposition}

\theoremstyle{definition}

\newcommand{\on}[1]{\operatorname{#1}}
\newcommand{\bs}[1]{\bm{#1}}
\newcommand{\Mmu}{\bs{M}_{\bs\mu}}
\newcommand{\Mrho}{\bs{M}_{\bs\rho}}
\newcommand{\hatMmu}{\widehat{\bs M}_{\bs\mu}}

\newcommand{\Span}{\on{span}} 
\newcommand{\Tr}{\on{Tr}}
\newcommand{\Var}{\on{Var}}
\newcommand{\MSE}{\on{MSE}}
\newcommand{\pinv}{\mathrm{pinv}}
\newcommand{\OT}{\mathrm{OT}}
\renewcommand{\sp}{\text{s-p}}
\newcommand{\mup}{{\mu\text{-p}}}
\newcommand{\nout}{n_{\mathrm{out}}}
\newcommand{\ntr}{n_{\mathrm{tr}}}
\newcommand{\es}[1]{\widehat{#1}}

\newcommand{\E}{\mathbb{E}}
\newcommand{\calD}{\mathcal{D}}
\newcommand{\calP}{\mathcal{P}}

\newcommand{\F}{\mathcal{F}}
\newcommand{\T}{\mathcal{T}}
\newcommand{\rmd}{\mathrm{d}}

\newcommand{\proj}{{\ket\phi\!\bra\phi}}
\newcommand{\Fte}{\overline{\F}_\mathrm{test}}
\newcommand{\G}{\mathcal{G}}
\newcommand{\dualmu}{\widetilde{\mu}}
\newcommand{\dualrho}{\widetilde{\rho}}
\newcommand{\Ob}{\mathcal{O}}
\newcommand{\rhotr}{{\bs\rho}}
\newcommand{\m}{{\hbn}}
\newcommand{\bmu}{{\bs\mu}}
\newcommand{\frameOpRho}{\es\F_\rho}

\newcommand{\norm}[1]{\left\lVert #1 \right\rVert}
\newcommand{\ket}[1]{\left\lvert #1 \right\rangle}
\newcommand{\bra}[1]{\left\langle #1 \right\rvert}
\newcommand{\braket}[2]{\left\langle #1 \middle| #2 \right\rangle}
\newcommand{\ketbra}[1]{\lvert #1\rangle\!\langle#1\rvert}

\newcommand{\RR}{\mathbb{R}}
\newcommand{\PN}{\widehat{\bP}_N}
\newcommand{\bw}{\bs w}
\newcommand{\bv}{\bs v}
\newcommand{\bp}{\bs p}
\newcommand{\bP}{\bs P}
\newcommand{\ex}{\mathrm{ex}}
\newcommand{\by}{\bs y}
\newcommand{\bSigma}{\bs\Sigma}
\newcommand{\hbp}{\widehat{\bp}}
\newcommand{\hbn}{\widehat{\bs n}}
\newcommand{\hw}{\widehat{\bw}}
\newcommand{\Cm}{\widehat{\bs C}_{\mu\rho}}

\NewCommandCopy{\rawcite}{\cite}
\NewCommandCopy{\rawcref}{\cref}
\NewDocumentCommand{\safecite}{om}{%
  \mbox{\IfValueTF{#1}{\rawcite[#1]{#2}}{\rawcite{#2}}}%
}
\NewDocumentCommand{\safecref}{m}{%
  \mbox{\rawcref{#1}}%
}
\NewDocumentCommand{\addLI}{+m +g}{%
  {\color{RoyalBlue}#1}%
  \IfNoValueF{#2}{%
    \begingroup
      \let\cite\safecite
      \let\cref\safecref
      \color{red}\small{#2}%
    \endgroup
  }%
}
\NewDocumentCommand{\addLIs}{+m +g}{%
  {\color{RoyalBlue}#1}%
  \IfNoValueF{#2}{%
    \begingroup
      \let\cite\safecite
      \let\cref\safecref
      \color{red}\small{\sout{#2}}%
    \endgroup
  }%
}
\newcommand{\rmvLI}[1]{{\color{red}\let\cite\safecite\let\cref\safecref\sout{#1}}}

\newcommand{\rmvGLM}[1]{{\color{red}\let\cite\safecite\let\cref\safecref\sout{#1}}}

\begin{document}
\pdfstringdefDisableCommands{\let\\\space}

\title{How to train quantum models with a single shot per state}

\newcommand{\unipa}{Universit\`a degli Studi di Palermo\comma{} Dipartimento di Fisica e Chimica - Emilio Segr\`e\comma{} via Archirafi 36\comma{} I-90123 Palermo\comma{} Italy}

\author{Alessandro Romancino\,\orcidlink{0009-0004-2812-6251}}
\email{alessandro.romancino@unipa.it}
\let\comma,
\affiliation{\unipa}

\author{Gabriele Lo Monaco\,\orcidlink{0000-0002-3594-3477}}
\let\comma,
\affiliation{\unipa}

\author{Salvatore Lorenzo\,\orcidlink{0000-0002-0827-5549}}
\let\comma,
\affiliation{\unipa}

\author{Alessandro Ferraro\,\orcidlink{0000-0002-7579-6336}}
\affiliation{Quantum Technology Lab\comma{} Dipartimento di Fisica Aldo Pontremoli\comma{} Universit\`a degli Studi di Milano\comma{} I-20133 Milano\comma{} Italy}

\author{Mauro Paternostro\,\orcidlink{0000-0001-8870-9134}}
\let\comma,
\affiliation{\unipa}
\affiliation{Centre for Quantum Materials and Technologies\comma{} School of Mathematics and Physics\comma{} Queen's University Belfast\comma{} BT7 1NN\comma{} United Kingdom}

\author{G. Massimo Palma\,\orcidlink{0000-0001-7009-4573}}
\let\comma,
\affiliation{\unipa}

\author{Luca Innocenti\,\orcidlink{0000-0002-7678-1128}}
\let\comma,
\affiliation{\unipa}

\date{\today}

\begin{abstract}\noindent
Quantum linear regression models, in which features obtained by measuring a fixed quantum device are combined through linear classical postprocessing, are a promising class of non-variational quantum machine learning protocols. These models are typically trained by least-squares regression, an approach recently shown to perform poorly in the undersampled regime. Here, we introduce a training method that extracts information from measurement data in a more efficient way, avoiding the stochastic ill-conditioning inherent to least squares, while also reducing the computational cost. Our method performs well even with a single measurement shot per training state and can therefore be run online, updating the weights sequentially as training data stream in. Crucially, while least-squares training saturates at a level that is set by the number of shots, here, we show that the error continues to improve as more training data are added. We also show that the computational cost of training, a major barrier to the scalability of these models, can be reduced further by exploiting additional prior information, such as symmetries of the training-state ensemble and of the measured observables, similarly to physics-informed neural networks.
\end{abstract}

\maketitle

\columntoc

\section{Introduction}
\label{sec:intro}

Quantum machine learning (QML) seeks to exploit quantum systems as computational resources for learning tasks involving classical or quantum data~\cite{2022cerezo,2017dunjko}. Variational quantum algorithms, in which parametrized quantum dynamics are optimized iteratively on training data, represent a widespread QML paradigm~\cite{2021cerezo}, but their optimization can suffer from barren plateaus and related trainability issues~\cite{2018mcclean,2020tangpanitanon,2021franca,2021wang,2022holmes,2022arrasmith,2023hu,2024gentinetta,2024thanasilp,2025larocca,2025cerezo,2025sweke,2025sahebi,2026kempkes}.
A large body of work~\cite{2018montanaro,2023anshu,2025banchi,2023gebhart,2019schuld} has focused instead on protocols based on fixed quantum dynamics, where learning is restricted to a linear classical postprocessing stage. This category includes, among others, quantum kernel methods~\cite{2019havlicek,2021kubler,2021schuld,2023jager,2023jerbi,2024thanasilp,2026kairon}, quantum reservoir computing (QRC)~\cite{2017fujii,2019ghosh,2020fujii,2021mujal}, quantum extreme learning machines (QELMs)~\cite{2023innocenti,2024lomonaco,2025vetrano,2025xiong,2025krisnanda,2023zia}, and related protocols for quantum property estimation and measurement postprocessing~\cite{2023tran,2024li,2024khan,2026zhang}.
Since these approaches share a common formal structure, and in particular rely on linear postprocessing of measurement data, we collectively refer to them as quantum linear regression (QLR) models. QLR approaches have shown promising results in a variety of applications, including photonic systems~\cite{2024suprano,2025zia,2026bartolo,2026brusaschi}, molecular physics~\cite{2024lomonaco}, and astrophysics~\cite{2025vetranoa}.

An unavoidable physical constraint of quantum learning protocols is that quantum features are accessible only through a finite number of measurements. The resulting sampling noise limits the information accessible from quantum features and increases the resources required for learning~\cite{2023hu,2024gentinetta,2024thanasilp,2026keller,2024ahmed,2025vetrano,2026meyer}. In two recent works~\cite{2026lomonaco,2026lomonacoa}, some of us initiated a systematic study of the effect of finite measurement statistics on QLR protocols, using measurement probabilities as quantum features.
In particular, it was shown in~\cite{2026lomonaco} that the standard least-squares training approach, namely fitting training weights via pseudo-inverse, becomes inefficient in the undersampled regime, where only a few measurement shots are available for each training state: in this regime empirical feature vectors are extremely noisy, causing prediction error to saturate even as the number of training states increases. This behavior runs counter to the classical intuition that larger training sets and resources should progressively improve the learned model~\cite{2017hestness,2022viering}. The fundamental question is then if this limitation is intrinsic to QLR or a consequence of how the measurement data are processed. Here, we show that it is the latter.

To answer this question, we introduce a novel training method, which we will refer to as \emph{Online Training} (OT), which overcomes the limitations of least-squares training. Our method works by first obtaining an effective description of the quantum measurement from the training states via a shadow-like construction~\cite{2018aaronson,2020huang,2022elben,2023innocentia,2023kunjummen,2024levy}, and then building the readout from the dual of the estimated measurement~\cite{2023innocenti}, in the spirit of task-oriented approaches to quantum process learning~\cite{2023huang}. As the name suggests, OT allows the readout to be updated sequentially as new training data become available, without the need to wait for the full training dataset to be measured. It is online in the sense of streaming estimators~\cite{2026marso}, rather than in the sense of learning from adaptively chosen data~\cite{2024raza}. Moreover, even though the procedure consists of two steps, we show that it is computationally more efficient than pseudo-inverse training and that we can take advantage of prior knowledge about global properties of the training set or measurement setup to make the training even more computationally efficient.

To showcase the improved performances of our method, on top of rigorous analytical scalings, we provide extensive numerical evidence showing the dramatically improved performances reachable via OT compared to standard least-squares training.
In particular, we show that with OT the MSE scales with the inverse of the total number of measurements, even using a single shot per state; by contrast, in this same regime least-squares regression is effectively unusable. We further show numerically that this new training method is immune to shot-noise-induced double descent phenomena, which are known to affect QLR training with least-squares~\cite{2026lomonacoa,2026meyer}.

OT borrows tools from process and detector tomography~\cite{2001fiurasek,2009lundeen,2009feito,2009bisio,2012zhang,2021wanga,2023xiao,2026zambrano,2018keith,1997chuang,1997poyatos,2013qi,2018guta,2022surawy-stepney,2014motka,2017motka} and puts them at the service of QLR training. From a statistical point of view, pseudo-inverse training is an errors-in-variables regression~\cite{1987fuller,2010buonaccorsi}, and OT is equivalent to its standard correction, two-stage least squares (2SLS), with the training states as instrumental variables~\cite{2001hausman,2010wooldridge} (\cref{app:iv}).

The rest of the article is organized as follows. In \cref{sec:linear-regression} we introduce QLR, standard least-squares training, and its finite-shot floor. In \cref{sec:ot} we introduce OT, show the advantages over least-squares, and study how both depend on the input dimension. In \cref{sec:physics-priors} we write the method in terms of frames and derive the prior variants of the method. We conclude with \cref{sec:conclusions}.

\section{Quantum linear regression}
\label{sec:linear-regression}

In quantum linear regression models (QLR), a fixed quantum device produces \textit{quantum features}, for example in the form of estimated probabilities or expectation values of a set of observables, or inner products between training states in the case of quantum kernels. Predictions are then computed as linear combinations of these features, and the training weights are learned solving the associated classical linear regression problem. QLR encompasses a range of non-variational quantum machine learning algorithms, such as QELMs, QRC, and quantum kernel methods, all sharing the same underlying formal structure~\cite{2020fujii,2021mujal,2019havlicek,2021schuld,2023innocenti,2026lomonaco, 2026lomonacoa,2021kubler}. Since each expectation value can only be estimated from a finite number of measurement shots, the features are intrinsically affected by shot noise. Throughout the article, we take QELMs processing quantum states as our reference model~\cite{2023innocenti}: the features are the outcome probabilities of a fixed measurement, that is, the expectation values of the effects of a POVM.

In this section, we review the main definitions and results for this setting, starting from the least-squares readout and its training (\cref{sec:qlr-readout}), and then its finite-shot error (\cref{sec:qlr-floor}).

\subsection{Least-squares readout of a fixed quantum device}
\label{sec:qlr-readout}

\parTitle{General setup}
In QLR, an input state $\rho$ on a $d$-dimensional Hilbert space $\mathcal H$ is processed by a fixed quantum channel $\mathcal E$, which we call the reservoir and which may include ancillary degrees of freedom and internal dynamics. The output of the quantum channel is then measured, and such a fixed physical measurement is modeled by a POVM  $\{\mu_a^\mathrm{phys}\}_{a=1}^{\nout}$ with $\nout$ outcomes. In the Heisenberg picture, channel and measurement together define the \emph{effective POVM} $\bmu=\{\mu_a\}_{a=1}^{\nout}$, with $\mu_a=\mathcal E^\dagger(\mu_a^\mathrm{phys})$, which acts directly on the input state~\cite{2023innocenti}. The quantum feature associated to a state $\rho$ is the probability vector $\bp(\rho)\equiv\langle\bs\mu,\rho\rangle\in\RR^{\nout}$, with elements $p_a(\rho)\equiv\langle\mu_a,\rho\rangle$. Here $a$ labels the possible measurement outcomes, and $\langle A,B\rangle=\Tr(A^\dagger B)$ is the Hilbert--Schmidt inner product. We use the shorthand notation $\langle\bs\mu,\rho\rangle$ to more expressively denote the vector of probabilities obtained measuring $\rho$ with the POVM $\bs\mu$. Collecting the outcome probabilities of a set of training states $\rhotr=\{\rho_i\}_{i=1}^{\ntr}$ we get the probability matrix $\bP$, defined so that $P_{ai}\equiv \langle\mu_a,\rho_i\rangle$, and written explicitly as
\begin{equation}
    \bP = \langle\bmu,\rhotr\rangle = \Mmu^T\Mrho,
    \label{eq:probability_matrix}
\end{equation}
where $\Mmu\equiv\sum_a \bv_{\mu_a}\bs e_a^T$ and $\Mrho\equiv\sum_i \bv_{\rho_i}\bs e_i^T$ are the matrices whose columns are the vectorized measurements and states, respectively. Here $\bv_A\in\RR^{d^2}$ collects the coordinates of a Hermitian operator $A$ in a fixed orthonormal basis of Hermitian operators, so that $\langle A,B\rangle=\bv_A^T\bv_B$, and $\bs e_a$ denotes the standard basis vectors~\cite{2026lomonaco,2026lomonacoa}. The particular structure of~\cref{eq:probability_matrix} has important implications for the overall performance of QLR and will be pivotal to our analysis of OT as well.

\parTitle{Training labels}
We focus on the task of predicting, for any input state $\rho$, the expectation value $y(\rho)=\langle\Ob,\rho\rangle$ of a target observable $\Ob$.
We consider a single target observable for simplicity. For $n_\mathrm{obs}$ observables, the readout $\bw$ and the training labels $\by$ introduced below become $\nout\times n_\mathrm{obs}$ and $\ntr\times n_\mathrm{obs}$ matrices, and all the results apply column by column.
Given any vector of $\nout$ weights $\bw$, the predicted expectation value of $\Ob$ on a test state $\sigma$ is computed from the corresponding quantum features as
\begin{equation}
    \es y(\sigma)=\bw^T\,\bp(\sigma)
    =\sum_{a=1}^{\nout} w_a\langle\mu_a,\sigma\rangle
    =\langle\Ob_{\bw},\sigma\rangle,
    \label{eq:effective-obs}
\end{equation}
where $\Ob_{\bw}=\sum_{a=1}^{\nout} w_a\mu_a$. Whatever training method produced $\bw$, a linear readout returns the expectation value of an associated \emph{effective} observable in $\Span\{\mu_a\}$. The goal of the training is thus to find a $\bw$ such that $\Ob_{\bw}\simeq\Ob$; in this sense, training amounts to an observable reconstruction problem~\cite{2023innocenti}. If the effective POVM is informationally complete (IC), its effects span the $d^2$-dimensional real space of Hermitian operators, and every observable can be reconstructed exactly, if enough training states are supplied.

\parTitle{Ideal training weights}
The standard method to find the weights $\bw$ is to solve the classical linear regression problem on the quantum features. Namely, given a training dataset $\{(\bp_i,y_i)\}_{i=1}^{\ntr}$, with $\bp_i\equiv\bp(\rho_i)$ the quantum features of the training state $\rho_i$ and $y_i=\langle\Ob,\rho_i\rangle$ its label, we look for the $\bw$ that minimizes the squared residual $\|\bP^T\bw - \by\|_2^2$. In the ideal limit of infinite statistics, where the quantum features are known exactly, and provided the training states span the space of Hermitian operators and $\Ob\in\Span\{\mu_a\}$, this returns the \emph{ideal readout}
\begin{equation}
    \bw_\ex \equiv \Mmu^+\bv_\Ob,
    \label{eq:ideal-readout}
\end{equation}
the minimum-norm solution of $\Ob_{\bw}=\Ob$.

\parTitle{Training with noisy features}
In practice, these features are not directly accessible, but must be experimentally determined by feeding each quantum state to the quantum device and measuring it. In the QELM reference model, each training state is measured $N$ times, producing a count vector $\hbn_i\sim\mathrm{Mult}(N,\bp_i)$, whose component $\es n_{ai}$ records how many times the outcome $a$ was observed. The empirical frequencies $\hbp_i=\hbn_i/N$ satisfy
\begin{equation}
    \E[\hbp_i\mid\rho_i]=\bp_i, \qquad
    \on{Cov}[\hbp_i\mid\rho_i]=\frac{\bSigma_i}{N},
    \label{eq:multinomial-moments}
\end{equation}
where $\bSigma_i\equiv\bSigma(\bp_i)$, and $\bSigma(\bp)\equiv\on{diag}(\bp)-\bp\bp^T$ is the categorical covariance matrix. The frequency vectors form the columns of the empirical probability matrix
\begin{equation}
    \PN\in\RR^{\nout\times\ntr}, \qquad
    (\PN)_{ai} = \frac{\es n_{ai}}{N},
    \label{eq:empirical-probs}
\end{equation}
whose average is the exact probability matrix $\bP$, with columns $\bp_i$.

Standard training fits the labels by least squares on the empirical features, $\PN^T\bw\simeq\by$. Its minimum-norm solution is given by the Moore--Penrose pseudo-inverse~\cite{1956penrose,2012barata},
\begin{equation}
    \hw_\pinv = (\PN^T)^+\by,
    \label{eq:pinv-readout}
\end{equation}
which is the standard way of training QELMs. Another least-squares approach that we will consider is ridge regression, that instead minimizes $\ntr^{-1}\|\PN^T\bw-\by\|_2^2+\tau\|\bw\|_2^2$, and gives $\hw_\tau=(\PN\PN^T+\ntr\tau I)^{-1}\PN\by$~\cite{2009hastie,2026lomonacoa}. At finite statistics, these readouts no longer return the ideal readout: sampling noise is known to introduce a systematic bias, so that $\E[\Ob_{\hw_\pinv}] \neq \Ob$, as we discuss next.

\begin{figure*}
    \centering
    \includegraphics[width=\textwidth]{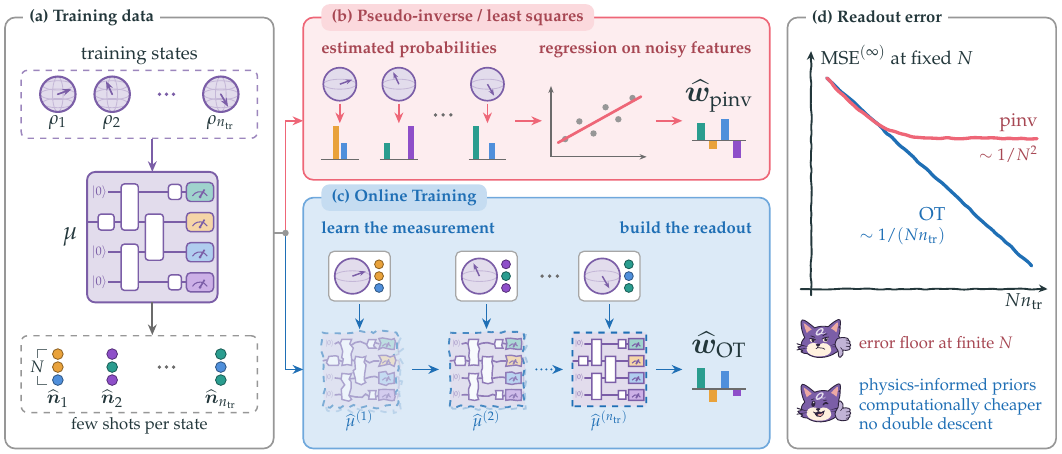}
    \caption{Pseudo-inverse training against Online Training.
    \emph{(a)} The training data: the training states $\rho_i$, the columns of $\Mrho$, are measured with the effective POVM $\bmu$ of the reservoir, with one effect $\mu_a$ per outcome. Each state is measured $N$ times, and its outcome counts $\hbn_i$ give the column $i$ of the empirical probability matrix $\PN$.
    \emph{(b)} Pseudo-inverse (and least-squares) training uses every state only through its few outcomes, as an estimate of its outcome probabilities, and fits the labels by regression on these noisy features, $\hw_\pinv=(\PN^T)^+\by$. Since $\E[\PN^+]\neq\bP^+$, the readout is biased and its error saturates at a finite-shot floor $b^2/N^2$.
    \emph{(c)} OT uses the pairs $(\rho_i,\hbn_i)$, one at a time, to learn the effective measurement: every pair refines the estimate $\widehat\mu^{(i)}$ of the measurement, which ends in the unbiased estimate $\hatMmu=(\Mrho^+)^T\PN^T$. The readout is then built from the learned measurement, $\hw_\OT=\hatMmu^+\bv_\Ob$.
    \emph{(d)} Schematic behavior of $\MSE^{(\infty)}$ as a function of the total number of measurement events $N\ntr$, at fixed $N$: the pseudo-inverse error stops at the finite-shot floor, while the OT error keeps decreasing as $1/(N\ntr)$. OT also admits physics-informed priors, is computationally cheaper, and avoids the double-descent peak of pseudo-inverse training.}
    \label{fig:ot-setup}
\end{figure*}

\subsection{Finite-shot error of least-squares training}
\label{sec:qlr-floor}

\parTitle{MSE at infinite test statistics}
To measure how good a trained readout $\hw$ is, we study its squared error averaged over the possible input states. Since we are interested in the error introduced during training, we evaluate the readout with exact test probabilities, which gives
\begin{equation}
    \MSE^{(\infty)}(\hw)
    \equiv \E_\sigma\!\left[\left(\es y(\sigma)-y(\sigma)\right)^2\right]
    = \E_\sigma\!\left[ \langle\Ob_{\hw}-\Ob,\sigma\rangle^2 \right],
    \label{eq:mse-inf}
\end{equation}
where the average is over the test states $\sigma$. Measuring each test state with a finite number of shots would simply add a test-sampling variance, which decreases with the number of test shots~\cite{2026lomonaco}.

\Cref{eq:mse-inf} is the error of one trained readout, which is itself random: it depends on the training data $\mathcal D_\mathrm{tr}=\{(\rho_i,\hbn_i)\}_{i=1}^{\ntr}$, that is, on which training states were drawn and on which outcomes were observed. Averaging over them gives
\begin{equation}
    \MSE^{(\infty)}(N,\ntr) \equiv \E_{\mathcal D_\mathrm{tr}}\!\left[\MSE^{(\infty)}(\hw)\right],
    \label{eq:mse-avg}
\end{equation}
a function of the training resources $N$ and $\ntr$ only. The asymptotic laws discussed in this article refer to this averaged error, which we simply write as $\MSE^{(\infty)}$ with a subscript for the training method.

\parTitle{Limitations of pseudo-inverse-based training strategies}
The performance of QLR is limited by several factors common to every training method: the number of training states, the expressivity of the effective POVM~\cite{2023innocenti,2025vetrano,2025lomonaco,2025xiong}, the conditioning of the exact feature matrix~\cite{2011vershynin,2013koltchinskii,2016christensen}, the alignment of the quantum representation with the task~\cite{2021kubler,2023jager}, and concentration phenomena~\cite{2022li,2026kairon,2026dao, 2026lomonacoa, 2026meyer}.
Here we focus instead on how the measurement data are processed during training, which matters most in the undersampled regime. Here, each column of $\PN$ is a poor estimate of the corresponding probability vector. Pseudo-inverse training puts this noise into a nonlinear and discontinuous function of the data~\cite{2026lomonaco}.

Since the Hermitian operators on $\mathcal H$ form a $d^2$-dimensional real space, $\on{rank}(\bP)\leq d^2$, so that in the overcomplete regime $\ntr,\nout>d^2$ the exact matrix is rank deficient. Its noisy estimate $\PN$, instead, is full rank with high probability: shot noise also fills the directions that the exact features never reach. Because the pseudo-inverse jumps when the rank changes~\cite{1977stewart,1990stewart}, $\PN^+$ does not converge to $\bP^+$ as $N\to\infty$, and $\E[\PN^+]\neq\bP^+$ even though $\E[\PN]=\bP$. In this case, the error of the trained readout treats the shot noise as part of the features, as in errors-in-variables regression~\cite{1995bishop,2012loh}, producing a bias that depends on the number of shots per state but not on the number of training states. Indeed, Ref.~\cite{2026lomonaco} showed that pseudo-inverse training gives
\begin{equation}
    \MSE_\pinv^{(\infty)} \simeq \frac{b^2}{N^2} + \frac{V}{N\ntr},
    \label{eq:mse-pinv}
\end{equation}
where $b^2$ and $V$ are the bias and variance coefficients averaged over the test states. Adding training states at fixed $N$ only reduces the variance, so the error saturates at the finite-shot floor $b^2/N^2$, and regularization, which acts on the same noisy features, does not remove it.

\section{Online Training}
\label{sec:ot}

The limitations of least-squares training  call for a training procedure that can better exploit the information in the measurement data, even in the severely undersampled regime. In this section, we introduce OT, a training method that provides an MSE scaling $\propto 1/(N\ntr)$ in all regimes, even where the measured features are extremely noisy. We first present the new method and its advantages over pseudo-inverse training (\cref{sec:main}), and then study how the error of both methods depends on the input dimension (\cref{sec:mse-dimension}).

\subsection{Online Training in a nutshell}
\label{sec:main}

\parTitle{Online training}
Pseudo-inverse training uses the training states only through the labels, which can be written explicitly in vectorized notation as $\by=\langle\bs\rho,\Ob\rangle=\Mrho^T\bv_\Ob$, fitting them directly with the noisy frequencies $\PN$. This results in the weights $\hw_{\pinv}$ as given in~\cref{eq:pinv-readout}.

By contrast, our proposed method computes the weights $\hw_{\OT}$ as follows:
\begin{equation}\label{eq:explicit_estimator_OT}
    (1)\,\,\hatMmu\equiv (\Mrho^+)^T\PN^T,
    \qquad
    (2)\,\,\hw_\OT \equiv \hatMmu^+\bv_\Ob.
\end{equation}
These two steps are understood operationally as follows:
\begin{enumerate}
    \item[$(1)$] The first stage is a form of detector tomography of the effective measurement, in which the training states play the role of the probe states~\cite{2017motka,2001fiurasek,2009lundeen} and the columns of $\hatMmu$ are the estimated effects. The randomness of $\hatMmu$ is due to shot noise fluctuations, and since $\E[\PN]=\bP$, we have
    \begin{equation}
        \E[\hatMmu^T]=\Mmu^T\Mrho\Mrho^+=\Mmu^T
    \end{equation}
    whenever the training states span the operator space, making $\hatMmu^T$ an unbiased estimator for the effective measurement.
    \item[$(2)$] The second stage builds the readout from the learned measurement exactly as the ideal readout of \cref{eq:ideal-readout} is built from the true one, as the minimum-norm solution of $\hatMmu\bw=\bv_\Ob$~\cite{2023innocenti}. In this step shot noise enters through $\hatMmu$, and because the optimal solution for $\bw$ is nonlinear in $\hatMmu$, the resulting readout $\hw_\OT$ is unbiased only to leading order.
\end{enumerate}
At inference, the two methods remain identical, and involve applying the computed $\hw$ vector to estimated probabilities on a new test state. We will show in \cref{sec:method} that this method also has a clear interpretation via the language of measurement frames, and in particular the first step is defining an unbiased estimator for the effective POVM that operates at the level of individual state-outcome pairs $(\rho,b)$.

\parTitle{Online vs pseudo-inverse training}
Using $\by = \Mrho^T\bv_\Ob$ in \cref{eq:pinv-readout}, the two training procedures can be written side by side as
\begin{equation}
\begin{aligned}
    \hw_\OT &= \big((\Mrho^+)^T\PN^T\big)^+\bv_\Ob,
    \\
    \hw_\pinv &= (\PN^T)^+\Mrho^T\bv_\Ob.
\end{aligned}
\label{eq:ot-pinv-matrix-comparison}
\end{equation}
Both readouts are computed from the same data, $\Mrho$ and $\PN$, and the difference lies only in how the matrices are inverted. Pseudo-inverse training inverts the noisy matrix $\PN$ directly, so that the shot noise of every training state enters the inversion and produces the finite-shot bias. OT instead inverts the known matrix $\Mrho$, so that the noisy frequencies enter $\hatMmu$ only linearly, and its second inversion acts on $\hatMmu$, whose fluctuations average out over all the measurement events and vanish as $\ntr\to\infty$ even at $N=1$. In the language of errors-in-variables regression, the known training states act as instruments for the noisy frequencies~\cite{2001hausman,2010wooldridge} as shown in \cref{app:iv}.

\begin{figure}[t]
    \centering
    \includegraphics[width=\columnwidth]{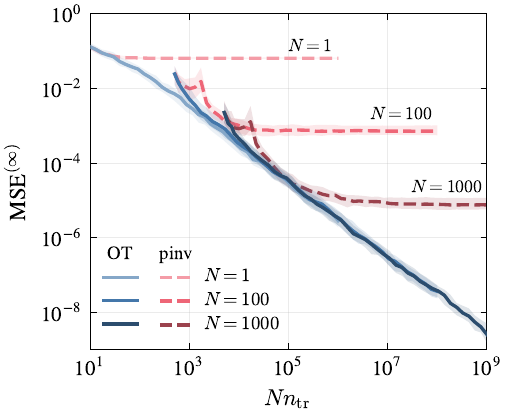}
    \caption{Scaling of $\MSE^{(\infty)}$ for OT and pseudo-inverse (pinv) as a function of the total resources $N\ntr$, for different values of $N$.
    While pseudo-inverse training (dashed red lines) saturates at $\sim1/N^2$, due to the bias term in~\cref{eq:mse-pinv}, with OT (solid blue lines) the MSE keeps improving as $1/(N\ntr)$ regardless of how the resources are partitioned between $N$ and $\ntr$.
    Here we show this numerically for single-qubit inputs and Haar POVMs with $\nout=16$ outcomes, for $N=1,100,1000$ shots per state (from light to dark).
    At large $N$, the pseudo-inverse error also shows a peak at $\ntr\simeq\nout$, one of the peaks of the triple-descent behavior of QLR~\cite{2026lomonacoa}, which is not present in the OT method.
    Ridge regression with a small regularization (not shown) gives the same curves as pseudo-inverse training, except that it removes this peak.
    Curves show the median over $100$ independent repetitions and shaded bands the 30th to 70th percentiles.}
    \label{fig:mse-ot-pinv}
\end{figure}

\parTitle{MSE scaling with OT}
The two methods in \cref{eq:ot-pinv-matrix-comparison} lead to a different dependence of the MSE on the training resources. For OT, when the training states span the operator space, the effective POVM is IC~\cite{2023innocenti}, and the estimation errors are small, we show in~\cref{app:perturbation} that the leading scaling of the MSE is given by
\begin{equation}
    \MSE_\OT^{(\infty)} \simeq \frac{V_\OT}{N\ntr}.
    \label{eq:mse-ot-main}
\end{equation}
The coefficient $V_\OT$ depends on the training ensemble, the measurement, and the target, but importantly not on $N$ and $\ntr$. Thus, to leading order, measuring a new training state or measuring again an old one reduces the error by the same amount. This is in stark contrast to the MSE scaling for pseudo-inverse training, which as shown in~\cref{eq:mse-pinv} has a bias term $b^2/N^2$, which causes performances to saturate while increasing $\ntr$, and for which $N$ and $\ntr$ play an asymmetric role.

\parTitle{Numerical results}
\Cref{fig:mse-ot-pinv} shows the difference in performance between OT and pseudo-inverse training, as measured via $\MSE^{(\infty)}$, as a function of the total resources $N\ntr$. For $\ntr >d^2$, the OT curves for different $N$ collapse onto a single line decreasing as $1/(N\ntr)$, as predicted by \cref{eq:mse-ot-main}, and only the total number of measurement events matters, even with a single shot per state. On the other hand, the pseudo-inverse curves saturate at the bias floor $b^2/N^2$, which decreases with $N$ but not with $\ntr$. Near the interpolation threshold $\ntr\simeq\nout$, the pseudo-inverse error also shows a double-descent peak~\cite{2026lomonaco,2026lomonacoa}, which OT avoids since it never inverts $\PN$. This data is obtained taking as target a rank-one projector $\Ob=\ketbra{\phi}$ with a Haar-random pure state $\ket\phi$, and Haar-random pure training and test states. The effective POVM is also resampled at each experiment realization as a Haar-random rank-1 POVM, that is, taking $\mu_a=\ketbra{k_a}$ with $\ket{k_a}\equiv K^\dagger \ket a$ the (unnormalized) row of a Haar-random isometry $K\in\mathbb C^{\nout\times d}$~\cite{2007mezzadri,2020heinosaari}.

\parTitle{What makes OT online}
An additional advantage of our method is that the estimator can be updated online as new data stream in. Since $\Mrho^+=\Mrho^T(\Mrho\Mrho^T)^+$~\cite{2012barata}, the estimate of \cref{eq:explicit_estimator_OT} factorizes as $\hatMmu^T=(\PN\Mrho^T)(\Mrho\Mrho^T)^+$, and both factors are sums with one term per training state whose size does not grow with $\ntr$: $\Mrho\Mrho^T$ is the frame operator of the training states, and the $a$-th row of $\PN\Mrho^T$ collects the training states weighted by how often they gave outcome $a$. Each new state and its outcomes can thus be added to the two sums and then discarded, and the readout is computed only when it is needed (\cref{alg:ot}). In \cref{sec:method} we show that this is the matrix form of an estimator that applies the dual frame of the training states to each outcome separately. At inference, OT is identical to standard training.

\parTitle{Computational cost of training}
Notably, the performance improvements brought forth by OT also come with an improved classical computational cost of training. Namely, we show in~\cref{app:complexity} that the cost of computing the readouts from training data in the two training approaches, in the overcomplete regimes we focus on, scales as
\begin{equation}
    T_\OT=O(\ntr\nout d^2), \qquad
    T_\pinv=O(\ntr\nout^2),
\end{equation}
which constitutes a gain of order $\nout/d^2$.

\subsection{Dependence on the input dimension}
\label{sec:mse-dimension}

\begin{figure*}
    \centering
    \includegraphics[width=\textwidth]{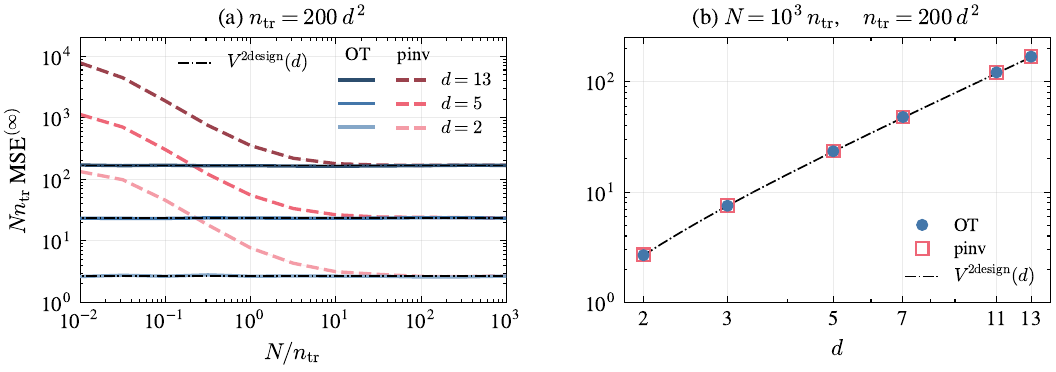}
    \caption{OT and pseudo-inverse (pinv) training reach the same error at high shots.
    Target $\proj$ and balanced 2-design POVM with $\nout=d(d+1)$ outcomes, built from a complete set of mutually unbiased bases~\cite{1989wootters,2005klappenecker} for prime $d$, $\ntr=\alpha_\rho d^2$ training states with $\alpha_\rho=200$.
    \emph{(a)} Rescaled error $N\ntr\MSE^{(\infty)}$ as a function of $N/\ntr$, for $d=2,5,13$ (from light to dark). The dash-dotted lines show $V^{\mathrm{2design}}(d)$.
    \emph{(b)} The same quantity at $N=10^3\ntr$ as a function of $d$. Since the coefficients are expectation values, lines and markers show arithmetic means, over $1000$ independent repetitions for $d\leq7$ and $100$ for $d=11,13$.}
    \label{fig:mse-dimension}
\end{figure*}

\parTitle{Variance scaling with dimension}
For any balanced POVM, whose effects all have the same trace $\Tr(\mu_a)=d/\nout$, the coefficients of the $1/(N\ntr)$ terms of the two methods coincide at large $\ntr$ and $V=V_\OT$. Thus OT fixes the bias floor intrinsic to the pseudo-inverse training, while at the same time not affecting its variance scaling $V/(N\ntr)$. If the POVM is also a 2-design, for Haar-random pure training and test states the common coefficient for the projector target is given by \cref{eq-app:2design-VOT},
\begin{equation}
    V^{\mathrm{2design}}(d) = \frac{d(d-1)(d+2)}{d+1},
    \label{eq:V-2design}
\end{equation}
while the bias coefficient is $b^2=(d-1)(d+2)^2/(d+1)$~\cite{2026lomonacoa}. Using \cref{eq:mse-ot-main,eq:mse-pinv} then gives
\begin{equation}
    \frac{\MSE_\pinv^{(\infty)}}{\MSE_\OT^{(\infty)}}
    \simeq 1 + \left(1+\frac{2}{d}\right)\frac{\ntr}{N},
    \label{eq:mse-ratio}
\end{equation}
so that, at every dimension, the pseudo-inverse reaches the OT error once $N\gg\ntr$.

\parTitle{Numerical tests of the dimension dependence of the MSE}
\Cref{fig:mse-dimension} tests these predictions with a fixed balanced 2-design POVM built from a complete set of mutually unbiased bases, while the training states and the target are redrawn at every repetition. In panel (a), the rescaled OT error $N\ntr\MSE_\OT^{(\infty)}$ lies on $V^{\mathrm{2design}}(d)$ for every $N$, even with only a few shots per state. The rescaled pseudo-inverse error is much larger for $N\ll\ntr$, where the bias term $b^2/N^2$ dominates, and joins the OT line once $N\gg\ntr$. In panel (b), at $N=10^3\ntr$, both methods coincide with $V^{\mathrm{2design}}(d)$ at every simulated dimension. Since $V^{\mathrm{2design}}$ grows as $d^2$, with $\ntr=\alpha_\rho d^2$ then the error $V^{\mathrm{2design}}/(N\ntr)\simeq1/(N\alpha_\rho)$ depends only weakly on $d$.

For POVMs whose effects have different traces, such as the Haar POVMs used in the rest of this work, $V\simeq V_\OT$ only in the limit $d\gg1$, where the effect traces concentrate around $d/\nout$: in our simulations the difference is about $20\%$ at $d=2$ and vanishes by $d=8$. It comes from the second stage of OT, whose readout is not the minimum-variance one for an unbalanced POVM, and it does not affect the absence of the floor.

\section{Physics-informed priors}
\label{sec:physics-priors}

In this section we present an alternative perspective on the OT method of~\cref{sec:ot}. We show that it arises naturally when looking for an unbiased estimator of the effective measurement that operates on individual state-outcome pairs. This explains why the method remains effective in the undersampled regime, and suggests how to make training even cheaper by injecting prior information on the distribution of the training states or on the structure of the effective POVM. We first derive OT from single shots (\cref{sec:method}), then write it in terms of frames for $N$ shots per state (\cref{sec:frames}), and finally introduce the prior-informed variants and their errors (\cref{sec:priors}).

\subsection{Single-shot formulation}
\label{sec:method}

\parTitle{Intuition via estimators}
To find a training strategy that keeps working even in the extremely undersampled regime, e.g.\ for $N=1$, we need a method that operates directly on individual state-outcome pairs $(\rho,b)$, where $b$ is the single outcome observed when measuring the training state $\rho$, so that both $\rho$ and $b$ are random. Pseudo-inverse training performs poorly in this regime, because it relies on estimated probability vectors, which for $N=1$ are too noisy to be useful.
We can, however, change perspective: we first look for an unbiased estimator of the effective measurement itself, and then use the estimated measurement to compute the readout weights.
To this end, observe that each measurement effect can be decomposed exactly as
\begin{equation}
    \mu_a = \int \rmd\calP(\rho)\,\sum_{b=1}^{\nout} \langle\mu_b,\rho\rangle \hat\mu_a(b,\rho),
    \label{eq:unbiased_povm_estimator}
\end{equation}
where $\rmd\calP(\rho)$ is the probability density of sampling the training state $\rho$ from some specified underlying ensemble, so that $\rmd\calP(\rho)\langle\mu_b,\rho\rangle$ is the joint probability of encountering the pair $(\rho,b)$ during training. Any function $\hat\mu_a(b,\rho)$ satisfying~\cref{eq:unbiased_povm_estimator} is thus an \textit{unbiased estimator} of the POVM element $\mu_a$: evaluating it at each measurement shot and averaging the results gives an estimate whose expectation value is exactly $\mu_a$. \Cref{eq:unbiased_povm_estimator} is solved by $\hat\mu_a(b,\rho)=\delta_{ab}\tilde\rho$, with $\tilde\rho\equiv \ntr\overline{\F}_\rho^{-1}(\rho)$, where the \textit{state frame operator}
\begin{equation}
    \overline{\F}_\rho(X)\equiv \ntr\int\rmd\calP(\rho)\, \langle X,\rho\rangle\rho
    \label{eq:population-state-frame}
\end{equation}
is the average frame of $\ntr$ states drawn from the ensemble. The operators $\tilde\rho$ are the \textit{canonical dual frame elements} of the training states, regarded as an operator frame~\cite{2006scott,2026innocenti}.

Once the POVM has been estimated, the readout is built with the standard shadow-estimation recipe for general POVMs, which relies on the dual of the POVM frame~\cite{2026innocenti}. If the effective POVM were known exactly, one would define the dual effects $\tilde\mu_a\equiv \F_\mu^{-1}(\mu_a)$, where
\begin{equation}
    \F_\mu(X)\equiv \sum_{a=1}^{\nout} \langle\mu_a,X\rangle\mu_a,
    \qquad
    \F_\mu=\Mmu\Mmu^T,
    \label{eq:true-povm-frame}
\end{equation}
is the POVM frame operator, so that $\hat o(a)\equiv \langle\Ob,\tilde\mu_a\rangle$ is an unbiased estimator of the expectation value of the target observable, $\E[\hat o\mid\rho]=\langle\Ob,\rho\rangle$. The corresponding weights $w_{\ex,a}=\langle\Ob,\tilde\mu_a\rangle$ are the components of the ideal readout of~\cref{eq:ideal-readout}. OT does precisely this, except that the unknown POVM is replaced by its estimate. Given the training records $\calD_{\rm tr}\equiv \{(\rho_i,b_i)\}_{i=1}^{\ntr}$, the resulting recipe is
\begin{equation}
    \es\mu_a \equiv \sum_{i\in I_a}\overline{\F}_\rho^{-1}(\rho_i),
    \qquad
    \es w_a \equiv \left\langle\Ob,\es\F_\mu^+(\es\mu_a)\right\rangle,
    \label{eq:intuitive-dual-states}
\end{equation}
where $I_a\equiv\{i:\,b_i=a\}$ is the set of indices of the training states that gave the outcome $a$, and $\es\F_\mu$ is the frame operator of the estimated effects. By linearity, $\es\mu_a=\overline{\F}_\rho^{-1}\big(\sum_{i\in I_a}\rho_i\big)$: each estimated effect is the dual of the training states that produced its outcome.
This recipe uses the frame operator $\overline{\F}_\rho$ of the ensemble from which the training states are drawn, and therefore requires knowing that ensemble: it is the state-prior variant of OT discussed in~\cref{sec:priors}. OT uses instead the frame operator of the training states actually measured, as we now show for a general number $N$ of shots per state.

\subsection{Multiple-shot formulation through frames}
\label{sec:frames}

The frame operator of the finitely many training states used in a given experiment is the \textit{empirical} state frame operator
\begin{equation}
    \frameOpRho(A)=\sum_{i=1}^{\ntr}\langle\rho_i,A\rangle\rho_i,
    \qquad
    \frameOpRho = \Mrho\Mrho^T,
    \label{eq:state-frame}
\end{equation}
where the second expression is its vectorized form. Its average over the training ensemble is the ensemble frame $\overline{\F}_\rho$ of~\cref{eq:population-state-frame}, and its support is the operator subspace spanned by the training states, so that it is invertible if and only if they span the operator space. Since $\frameOpRho(\mu_a)=\sum_i p_a(\rho_i)\rho_i$, applying $\frameOpRho^+$ gives, for spanning training states,
\begin{equation}
    \mu_a = \sum_{i=1}^{\ntr} p_a(\rho_i)\,\dualrho_i,
    \qquad
    \dualrho_i \equiv \frameOpRho^+(\rho_i),
    \label{eq:dual-states}
\end{equation}
the finite-ensemble counterpart of~\cref{eq:unbiased_povm_estimator}: the dual states $\dualrho_i$ reconstruct each effect from the exact outcome probabilities. OT replaces these probabilities with the observed frequencies $\es p_{ai}=\es n_{ai}/N$, which gives the two stages of~\cref{eq:explicit_estimator_OT} in frame form:
\begin{enumerate}
    \item[$(1)$] The estimated effects are
    \begin{equation}
        \es\mu_a = \frac{1}{N} \sum_{i=1}^{\ntr} \es n_{ai}\,\frameOpRho^+(\rho_i)
        = \sum_{i=1}^{\ntr} \es p_{ai}\,\dualrho_i ,
        \label{eq:povm-estimator}
    \end{equation}
    which for $N=1$ is~\cref{eq:intuitive-dual-states} with the empirical frame $\frameOpRho$ in place of the ensemble frame $\overline{\F}_\rho$. Since $\E[\es n_{ai}\mid\rho_i]=Np_a(\rho_i)$, \cref{eq:dual-states} gives $\E_{\m}[\es\mu_a\mid\rhotr]=\mu_a$ whenever the training states span the operator space: at fixed training states, the estimator is unbiased with respect to the measurement statistics $\m=\{\hbn_i\}_{i=1}^{\ntr}$.
    \item[$(2)$] The readout represents the target observable with the learned effects, using the frame operator of the estimated POVM~\cite{2023innocentia}, obtained from~\cref{eq:true-povm-frame} with $\mu_a\to\es\mu_a$,
    \begin{equation}
        \es\F_\mu(A) = \sum_{a=1}^{\nout} \langle\es\mu_a,A\rangle\es\mu_a,
        \qquad
        \es\F_\mu=\hatMmu\hatMmu^T,
        \label{eq:povm-frame}
    \end{equation}
    and its canonical dual $\dualmu_a \equiv \es\F_\mu^+(\es\mu_a)$, which give the weights
    \begin{equation}
        \es w_a = \langle \dualmu_a,\Ob\rangle
        = \left\langle \es\F_\mu^+(\es\mu_a), \Ob \right\rangle.
        \label{eq:ot-weights}
    \end{equation}
    Since $\es\F_\mu\es\F_\mu^+$ is the orthogonal projector onto $\Span\{\es\mu_a\}$, these weights reconstruct the target exactly within the estimated measurement model, $\sum_a\es w_a\es\mu_a=\Ob$, whenever $\Ob\in\Span\{\es\mu_a\}$. For overcomplete measurements other duals could be chosen to reduce the variance for specific observables or input ensembles~\cite{2024fischer,2025mangini}; here we use the canonical one, which is fixed by the learned measurement.
\end{enumerate}
The equivalence between these frame expressions and the matrix formulation of~\cref{eq:explicit_estimator_OT} is shown in~\cref{app:complexity}.

\subsection{Prior-informed OT}
\label{sec:priors}

The two empirical inverses of OT, $\frameOpRho^+$ to estimate the measurement and $\es\F_\mu^+$ to build its dual readout, act on $d^2\times d^2$ operators: constructing them costs $O(\ntr d^4)$ and $O(\nout d^4)$ operations, and each pseudo-inverse costs $O(d^6)$. When the ensemble from which the training states or the measurement are generated is known \emph{a priori}, the corresponding frame can replace the empirical one,
\begin{equation}
    \frameOpRho^+ \longrightarrow \overline{\F}_\rho^{-1}, \qquad
    \es\F_\mu^+ \longrightarrow \overline{\F}_\mu^{-1},
    \label{eq:prior-replacements}
\end{equation}
which avoids its construction and inversion, at the price of no longer adapting to the particular finite realization of the experiment.
Throughout, we assume that the priors are correct, that is, each population frame equals the ensemble average of the corresponding true frame. For training states sampled independently from the same ensemble, the ensemble frame of~\cref{eq:population-state-frame} is $\overline{\F}_\rho=\E_\rho[\frameOpRho]=\ntr\E_\rho[\bv_\rho\bv_\rho^T]$, which grows linearly with $\ntr$. For the measurement, $\overline{\F}_\mu\equiv\E_\mu[\F_\mu]$ is the average over the measurement ensemble of the frame $\F_\mu$ of the true effective POVM, \cref{eq:true-povm-frame}, given in \cref{eq:haar-povm-prior} for Haar POVMs; its scale is fixed by the completeness relation $\sum_a\mu_a=I$.
The state prior therefore simplifies the first reconstruction step using information about the preparation ensemble, whereas the POVM prior simplifies the second using information about the measurement ensemble. The latter assumption is more restrictive, since it requires statistical knowledge of the reservoir measurement and partly weakens the black-box character of the original regression problem~\cite{2024suprano,2026bartolo}.

\parTitle{State prior}
The ensemble of training states is often known independently of the reservoir, since it is fixed by the preparation procedure. For Haar-random pure states $\rho_\psi=\ket\psi\!\bra\psi$, the population state frame is~\cite{2013harrow,2024mele}
\begin{equation}
    \overline{\F}_\rho(A) = \ntr\int\rmd\psi\,\Tr(\rho_\psi A)\rho_\psi
    = \ntr\,\frac{\Tr(A)I+A}{d(d+1)},
    \label{eq:haar-state-frame}
\end{equation}
with inverse $\overline{\F}_\rho^{-1}(A) = [d(d+1)A - d\Tr(A)I]/\ntr$. Replacing the empirical inverse in \cref{eq:povm-estimator} gives~\cite{2026zambrano}
\begin{equation}
    \es\mu_a^\sp=\frac{1}{N}\sum_{i=1}^{\ntr}\es n_{ai}\,\overline{\F}_\rho^{-1}(\rho_i),
\end{equation}
and we call the resulting method \emph{state-prior OT}. Its readout still uses the empirical POVM-frame inverse, so that the measurement and its dual are still learned from the observed data.

\parTitle{POVM prior}
A prior for the effective measurement requires an additional physical assumption, since the POVM is not fixed by the preparation procedure but must come from a model of the reservoir dynamics and measurement statistics. It requires knowledge of the measurement ensemble, not of the effects of the realized POVM. For the Haar POVMs, the average POVM frame is~\cite{2020heinosaari,2024mele}
\begin{equation}
    \overline{\F}_\mu(A) = \E_\mu \left[\sum_{a=1}^{\nout} \langle\mu_a,A\rangle\mu_a\right]
    = \frac{\Tr(A)I + A}{\nout + 1},
    \label{eq:haar-povm-prior}
\end{equation}
with inverse
\begin{equation}
    \overline{\F}_\mu^{-1}(A) = (\nout + 1)A - \frac{\nout+1}{d+1}\Tr(A)I,
    \label{eq:haar-povm-prior-inv}
\end{equation}
and the readout weights become $\es w_a = \langle \es\mu_a, \overline{\F}_\mu^{-1}(\Ob) \rangle$. We call this variant \emph{POVM-prior OT} when the effects $\es\mu_a$ are obtained with the empirical state-frame inverse, and \emph{double-prior OT} when both substitutions of \cref{eq:prior-replacements} are made.

\parTitle{Error of the prior-informed variants}
The computational savings of the priors come with identifiable statistical costs, which we quantify with the training-averaged error of \cref{eq:mse-avg}. For double-prior OT, we show in \cref{app:analytical} that it decomposes exactly as
\begin{equation}
    \MSE^{(\infty)}(N,\ntr)=\beta^2+\frac{V_\OT}{N\ntr}+\frac{V_\rho}{\ntr}.
    \label{eq:bias-coefficients}
\end{equation}

\begin{figure}[tb]
    \centering
    \includegraphics[width=\columnwidth]{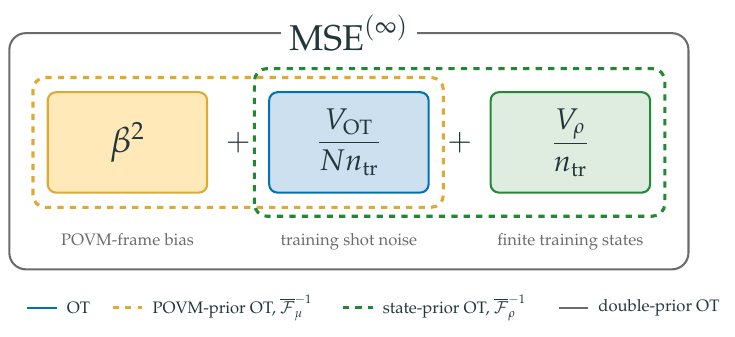}
    \caption{Error terms of OT and its prior-informed variants, \cref{eq:bias-coefficients}.
    OT retains only the shot-noise term $V_\OT/(N\ntr)$; the POVM prior adds the bias $\beta^2$ (yellow), the state prior adds the finite-state term $V_\rho/\ntr$ (green), and double-prior OT retains all three (gray).}
    \label{fig:ot-variants}
\end{figure}

\begin{itemize}
    \item The contribution $V_\OT/(N\ntr)$ is the training shot noise already encountered for OT. For fixed training states, the observed counts fluctuate according to multinomial statistics. Averaging $N$ measurements per state produces the factor $1/N$, while averaging over $\ntr$ independent training records gives the additional factor $1/\ntr$.
    \item The term $V_\rho/\ntr$ is specific to the state-prior replacement. The population inverse $\overline{\F}_\rho^{-1}$ reconstructs the measurement correctly after averaging over the preparation ensemble, but does not exactly cancel the fluctuations of a particular finite training set. The resulting finite-state contribution survives even when $N\rightarrow\infty$ and is instead suppressed by increasing $\ntr$. OT avoids this term because its empirical inverse is constructed from the same realization of the training states, as shown in \cref{app:perturbation}.
    \item Finally, $\beta^2$ is the systematic error associated with the POVM prior. Even when all statistical fluctuations vanish, the prior readout, with weights $\overline{w}_a=\langle\mu_a,\overline{\F}_\mu^{-1}(\Ob)\rangle$, reconstructs $\Ob_{\overline{\bw}} = \F_\mu\overline{\F}_\mu^{-1}(\Ob) \neq \Ob$. The asymptotic error is then
    \begin{equation}
        \beta^2 = \E_\sigma\!\left[\langle\Ob-\Ob_{\overline{\bw}},\sigma\rangle^2\right].
    \end{equation}
    Unlike the other terms, $\beta^2$ is independent of both $N$ and $\ntr$. It vanishes only when the prior measurement frame correctly represents the relevant action of the realized measurement on the target observable.
\end{itemize}

Applying the two substitutions of \cref{eq:prior-replacements} separately or together gives the four variants summarized in \cref{fig:ot-variants}, each with the error terms it retains, and \cref{app:priors} studies how the excess errors of the two priors decrease with the number of training states and of outcomes. The coefficients are method dependent; in particular, the $V_\OT$ coefficients of OT and state-prior OT coincide to leading order in the large-$\ntr$ population-frame limit (\cref{app:perturbation-sp}).

\begin{figure}[t]
    \centering
    \includegraphics[width=\columnwidth]{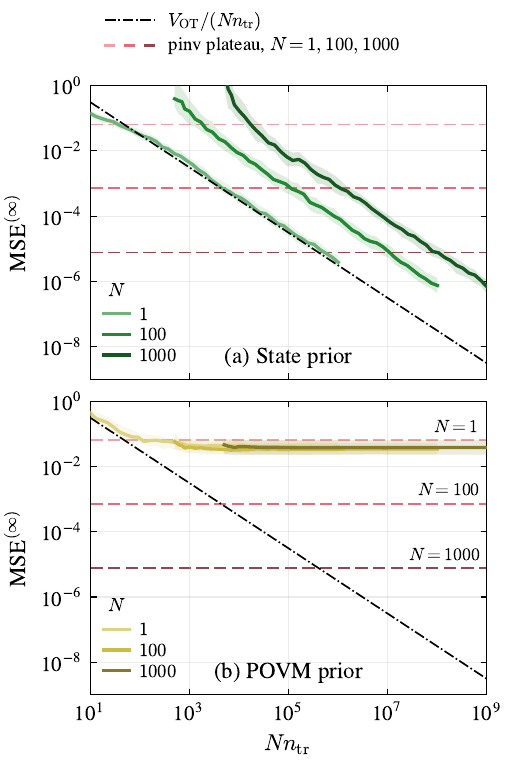}
    \caption{Prior-informed variants of OT, for a single-qubit input ($d=2$) with $\nout=16$ outcomes.
    $\MSE^{(\infty)}$ as a function of the total number of measurement events $N\ntr$, for $N=1,100,1000$ shots per state (from light to dark) and $\ntr>d^2$: \emph{(a)} state-prior OT, \emph{(b)} POVM-prior OT.
    The dash-dotted line is the OT law $V_\OT/(N\ntr)$, with $V_\OT$ fitted to the OT curves of \cref{fig:mse-ot-pinv}, and the dashed lines are the finite-shot floors of pseudo-inverse (pinv) training at the same values of $N$ (from light to dark). Double-prior OT is not shown, as it saturates at the same $\beta^2$ as POVM-prior OT.
    Curves show the median over $100$ independent repetitions and shaded bands the 30th to 70th percentiles.}
    \label{fig:mse-simulations}
\end{figure}

\parTitle{Numerical comparison}
\Cref{fig:mse-simulations} compares the two single-prior variants with the references of \cref{fig:mse-ot-pinv}, on the task of \cref{sec:main}. State-prior OT approaches the OT law only in the few-shot regime, since its finite-state term $V_\rho/\ntr$ does not decrease with $N$. Unlike pseudo-inverse training, however, it has no finite-shot floor, and at every $N$ its error eventually falls below the pseudo-inverse floor as $\ntr$ grows. POVM-prior OT instead saturates at $\beta^2$, which here lies just below the single-shot pseudo-inverse floor.

\section{Conclusions}
\label{sec:conclusions}

The standard approach to training QLR models, namely solving the classical least-squares problem on the measured noisy features, faces severe limitations in the undersampled regime, where the features are too noisy for pseudo-inverse-based methods to operate effectively~\cite{2026lomonaco}. In this work, we introduced OT, a novel training approach tailored to quantum measurement data, which performs well even in this undersampled regime, down to the case where each state is measured only once.

We showed in~\cref{sec:ot} that, while operating in exactly the same experimental setting and on the same data, this method has a lower computational cost and overcomes the saturation behavior typical of standard alternative approaches. Furthermore, it exhibits the same $\propto 1/(N\ntr)$ variance scaling as the pseudo-inverse approach~\cite{2026lomonaco}, while completely removing the term in the MSE responsible for the latter's saturation.
Our OT approach can also be deployed \textit{online}: the estimator can be accumulated iteratively as training data stream in, without storing the full training history, and the readout can then be constructed whenever needed. This contrasts with pseudo-inverse-based strategies, which require recomputing the estimated probability matrix whenever new training data are added, and stop working if one simply adds more undersampled columns to it.

We further showed in~\cref{sec:physics-priors} that OT arises naturally by borrowing ideas from shadow tomography on general frames~\cite{2023innocentia,2026innocenti}: one first defines an unbiased estimator of the effective measurement based on single measurement records, and then uses this estimate to construct a corresponding estimator for the target observable. Interestingly, when adapted to ensembles comprising finitely many training states, this general construction takes a form closely resembling that of pseudo-inverse training, except that the matrix inversions are performed in a different order. This same perspective also makes it natural to incorporate prior information into the training strategy, in the form of assumptions on the properties of the training-state ensemble and of the effective measurement. These priors allow the training to be performed at a lower computational cost, at the expense of introducing some bias into the associated estimator.

Our analysis of the dependence on the input dimension provides a further perspective on the leading error coefficient $V_\OT$. Under the Haar-averaged assumptions considered, $V_\OT$ is proportional to the average single-shot variance of the ideal canonical readout (\cref{eq-app:VOT-haar-integral}). Moreover, our numerical results show that OT and pseudo-inverse training approach the same leading coefficient for large $N$, despite being constructed from very different training procedures. This suggests that $V_\OT$ may capture a more general limit rather than a feature specific to OT. An interesting open question is whether, for IC measurements and in the absence of additional task-specific prior information, $V_\OT/(N\ntr)$ represents the minimal asymptotic MSE achievable by an unbiased QLR readout.

One possible route to address this question is to optimize the choice of the dual readout. The canonical dual used here is not generally the minimum-variance choice, and alternative dual constructions can reduce the fluctuations of observable estimators~\cite{2024fischer,2025mangini}. More generally, this suggests that the different experimental resources entering the learning problem could be optimized together.

Another interesting direction is to connect OT with Gaussian-process approaches to quantum learning~\cite{2025garcia-martin,2026jager}. Prior knowledge about the effective measurement could then be used not only to construct the readout, but also to quantify the uncertainty associated with its predictions. This uncertainty could help identify which additional training states would be most informative, providing a natural route towards adaptive online training.

Finally, a particularly natural extension of OT is towards temporal quantum reservoir computing. In this setting, the reservoir processes a continuous sequence of inputs and exploits its internal memory to solve time-dependent tasks. The online structure of OT could then allow the readout to be updated continuously as new data are processed, without storing the full sequence of measured features. An important question is whether the same statistical advantages persist when consecutive reservoir states are correlated and depend on the previous input history. Extending OT to this regime could open the way to genuinely online quantum learning for temporal tasks, and provide a promising setting in which to search for experimentally accessible quantum advantages.

\section*{Data availability}
The code supporting the findings, as a Python library where you can tinker with the OT method, is available on GitHub at \url{https://github.com/alex180500/online-qml}.

\section*{Acknowledgements}
GLM acknowledges funding from the European Union--NextGenerationEU through the Italian Ministry of University and Research under PNRR-M4C2-I1.3 Project PE-00000019 ``HEAL ITALIA'' (CUP B73C22001250006).
SL, AF, and GMP acknowledge support by MUR under PRIN Project No.~2022FEXLYB, Quantum Reservoir Computing (QuReCo).
LI, SL, MP, and GMP acknowledge funding from the ``National Centre for HPC, Big Data and Quantum Computing (HPC)'' Project CN00000013, HyQELM--Spoke 10.
MP is grateful for support from the Royal Society Wolfson Fellowship (RSWF/R3/183013), the Department for the Economy of Northern Ireland under the US-Ireland R\&D Partnership Programme, the PNRR PE Italian National Quantum Science and Technology Institute (PE0000023), and the EU Horizon Europe EIC Pathfinder project QuCoM (GA no.~10032223).
We acknowledge CINECA for the availability of high-performance computing resources and support on the Leonardo supercomputer, under the project \emph{qurachao}.

\bibliography{references}
\clearpage

\clearpage
\onecolumngrid
\appendix
\crefalias{section}{appendix}
\crefalias{subsection}{subappendix}

\setcounter{equation}{0}
\setcounter{figure}{0}
\setcounter{table}{0}
\setcounter{section}{0}

\makeatletter
\renewcommand{\thesection}{\Alph{section}}
\renewcommand{\theequation}{\thesection\arabic{equation}}
\renewcommand{\thefigure}{\thesection\arabic{figure}}
\renewcommand{\thetable}{\thesection\arabic{table}}
\@addtoreset{table}{section}
\@addtoreset{figure}{section}
\@addtoreset{theorem}{section}
\@addtoreset{lemma}{section}
\@addtoreset{proposition}{section}
\@addtoreset{corollary}{section}
\@addtoreset{conjecture}{section}
\@addtoreset{result}{section}
\@addtoreset{remark}{section}
\renewcommand{\thetheorem}{\thesection\arabic{theorem}}
\renewcommand{\thelemma}{\thesection\arabic{lemma}}
\renewcommand{\theproposition}{\thesection\arabic{proposition}}
\renewcommand{\thecorollary}{\thesection\arabic{corollary}}
\renewcommand{\theconjecture}{\thesection\arabic{conjecture}}
\renewcommand{\theresult}{\thesection\arabic{result}}
\renewcommand{\theremark}{\thesection\arabic{remark}}

\renewcommand{\theHsection}{SM.\Alph{section}}
\renewcommand{\theHequation}{SM.\Alph{section}.\arabic{equation}}
\renewcommand{\theHfigure}{SM.\Alph{section}.\arabic{figure}}
\renewcommand{\theHtable}{SM.\Alph{section}.\arabic{table}}

\renewcommand{\theHtheorem}{SM.\Alph{section}.\arabic{theorem}}
\renewcommand{\theHlemma}{SM.\Alph{section}.\arabic{lemma}}
\renewcommand{\theHproposition}{SM.\Alph{section}.\arabic{proposition}}
\renewcommand{\theHcorollary}{SM.\Alph{section}.\arabic{corollary}}
\renewcommand{\theHconjecture}{SM.\Alph{section}.\arabic{conjecture}}
\renewcommand{\theHresult}{SM.\Alph{section}.\arabic{result}}
\renewcommand{\theHremark}{SM.\Alph{section}.\arabic{remark}}
\makeatother

\makeatletter
\let\SMoriginaladdtocontents\addtocontents
\long\def\addtocontents#1#2{%
  \def\SMtocfile{#1}%
  \def\SMstandardtoc{toc}%
  \ifx\SMtocfile\SMstandardtoc
    \SMoriginaladdtocontents{smt}{#2}%
  \else
    \SMoriginaladdtocontents{#1}{#2}%
  \fi
}
\newcommand{\SMtableofcontents}{%
  \begingroup
    \small
    \setlength{\parskip}{0pt}%
    \def\l@f@section{\addvspace{0.2em}}%
    \let\l@subsubsection\@gobbletwo
    \let\SM@numberline\numberline@@sections
    \def\numberline@@sections##1{%
      \if\relax\detokenize{##1}\relax
        \leavevmode\ignorespaces
      \else
        \SM@numberline{##1}%
      \fi}%
    \@starttoc{smt}%
  \endgroup
}
\newcommand{\SMtwocoltableofcontents}{%
  \begingroup
    \small
    \setbox\z@\vbox{%
      \hsize=\dimexpr(\textwidth-1cm)/2\relax 
      \linewidth\hsize \columnwidth\hsize
      \interlinepenalty\@M \clubpenalty\@M \widowpenalty\@M 
      \vbox{}
      \SMtableofcontents}%
    \splittopskip\baselineskip \vbadness\@M \vfuzz\maxdimen
    \setbox\tw@\vsplit\z@ to\z@ 
    \dimen@=.5\ht\z@
    \loop 
      \setbox\tw@\copy\z@
      \setbox4\vsplit\tw@ to\dimen@
      \ifdim\ht\tw@>\dimen@ \advance\dimen@ 1pt \repeat
    \par\nointerlineskip
    \noindent\hbox to\textwidth{\vtop{\unvbox4}\hfil\vtop{\unvbox\tw@}}\par
  \endgroup
}
\makeatother

\makeatletter
\begingroup
  \let\addtocontents\SMoriginaladdtocontents
  \def\toclevel@section{0}%
  \phantomsection
  \addcontentsline{toc}{section}{Supplemental Material}%
\endgroup
\makeatother

\begin{center}
    {\Large\bfseries Supplemental Material}
\end{center}
\vspace{1em}
\noindent\textbf{Contents}\par\smallskip
\SMtwocoltableofcontents
\bigskip

\section{Perturbative expansion of OT}
\label{app:perturbation}

When both frame inverses are fixed before observing the training data, as in double-prior OT, the MSE decomposes exactly (\cref{app:analytical}). Here we consider OT and state-prior OT, which instead construct the readout from the estimated POVM frame. The inverse of this frame depends nonlinearly on the complete training dataset, so that the independent-contribution calculation of \cref{app:analytical} cannot be applied directly.

We first linearize the reconstructed observable in the POVM-estimation error in \cref{app:perturbation-linearization}. We then derive the leading MSE for OT in \cref{app:perturbation-ot} and identify the additional finite-state contribution of state-prior OT in \cref{app:perturbation-sp}. Finally, we evaluate the coefficients for Haar-random training and test states, comparing balanced 2-design measurements with the large-outcome limit of Haar POVMs in \cref{app:perturbation-benchmarks}.

As in \cref{eq:mse-avg}, all MSEs below are averaged over training realizations, and the superscript $(\infty)$ denotes exact test probabilities. The coefficients have the same interpretation as in \cref{eq:bias-coefficients}, but their values depend on the estimator. Throughout this appendix, the true POVM is fixed and IC. We assume that the empirical state frame is invertible and work in the regime where the estimated POVM frame remains well conditioned. The training states are independent samples from their ensemble, with independent measurement records, and the test states are independent of the training data.

\subsection{Linearization of the empirical readout}
\label{app:perturbation-linearization}

Following Ref.~\cite{2026lomonaco}, we write the empirical training frequencies in terms of the rescaled shot noise $\es{\bs\xi}_i\equiv\sqrt N(\hbp_i-\bp_i)$, with components $\es\xi_{ai}$,
\begin{equation}
    \hbp_i = \bp_i+\frac{\es{\bs\xi}_i}{\sqrt N}, \qquad
    \PN = \bP + \frac{\es{\bs\Xi}}{\sqrt N},
    \label{eq-app:perturbative-probs}
\end{equation}
where $\es{\bs\Xi}=(\es{\bs\xi}_1\,\cdots\,\es{\bs\xi}_{\ntr})$. From \cref{eq:multinomial-moments}, the multinomial statistics give
\begin{equation}
    \E_{\hbn_i}[\es{\bs\xi}_i\mid\rho_i] = 0,\qquad
    \E_{\hbn_i}[\es{\bs\xi}_i\es{\bs\xi}_i^T\mid\rho_i] = \bSigma_i,
    \label{eq-app:conditional-error}
\end{equation}
with $\bSigma_i=\on{diag}(\bp_i)-\bp_i\bp_i^T$. The rescaling makes the noise moments independent of $N$, so that all the $N$ dependence is explicit in the prefactors below.
Fluctuations associated with different training states are independent conditional on the training set $\rhotr$.

Let $\Mmu\in\RR^{d^2\times\nout}$ have the vectorized $\mu_a$ as its columns, as in \cref{sec:qlr-readout}, so that $\F_\mu=\Mmu\Mmu^T$. For either estimator we write $\es\mu_a=\mu_a+\delta\mu_a$ and $\hatMmu=\Mmu+\delta\Mmu$. The ideal readout $\bw_\ex$ of \cref{eq:ideal-readout} has components
\begin{equation}
    w_{\ex,a} = \left\langle\mu_a,\F_\mu^{-1}(\Ob)\right\rangle.
    \label{eq-app:ideal-dual-weights}
\end{equation}
The following result justifies retaining the first-order observable error when computing the MSE through second order.

\begin{proposition}[Linearized readout and estimation-error bound]
    \label{prop:ot-linearization}
    For a sufficiently small perturbation, $\norm{\delta\Mmu}_{\mathrm{op}} \ll \sqrt{\lambda_{\min}(\F_\mu)}$, the observable reconstructed by applying the empirical readout to the true POVM obeys
    \begin{equation}
        \delta\Ob_{\hw} \equiv \Ob_{\hw} - \Ob
        = \delta\Ob_{\hw}^{(1)}+O\!\left(\norm{\delta\Mmu}_F^2\right),
        \label{eq-app:delta-oeff}
    \end{equation}
    where we have defined $\delta\Ob_{\hw}^{(1)} \equiv -\sum_a w_{\ex,a}\delta\mu_a$. Consequently, for a test frame $\Fte$, its squared prediction error is
    \begin{equation}
        \langle\delta\Ob_{\hw},\Fte(\delta\Ob_{\hw})\rangle
        = \left\langle\delta\Ob_{\hw}^{(1)},\Fte(\delta\Ob_{\hw}^{(1)})\right\rangle
        + O\!\left(\norm{\delta\Mmu}_F^3\right).
        \label{eq-app:mse-quadratic}
    \end{equation}
    For OT, the conditional mean-square POVM error additionally satisfies
    \begin{equation}
        \E_\m\!\left[\norm{\delta\Mmu}_F^2 \middle| \rhotr\right]
        \leq \frac{\Tr(\es\F_\rho^+)}{N}.
        \label{eq-app:povm-error-bound}
    \end{equation}
    The constants in the remainders depend on the true frame and the target observable.
\end{proposition}

\begin{proof}
    The weights of \cref{eq:ot-weights} reconstruct the target within the learned measurement model $\sum_a\es w_a\es\mu_a=\Ob$. Their application to the actual reservoir instead gives $\Ob_{\hw}=\sum_a\es w_a\mu_a$. Subtracting the two expressions yields the exact identity $\delta\Ob_{\hw}=-\sum_a\es w_a\delta\mu_a$. Since the inverse frame is a smooth function of a full-rank frame under sufficiently small perturbations~\cite{1973wedin,2019xu}, we have the first-order expansion $\es w_a = w_{\ex,a} + O(\norm{\delta\Mmu}_F)$. Inserting this into the previous expression directly gives \cref{eq-app:delta-oeff}. Its first-order term is linear in $\delta\Mmu$, so the cross term with the second-order remainder is cubic. This gives \cref{eq-app:mse-quadratic}.
    
    To prove \cref{eq-app:povm-error-bound}, substitute \cref{eq-app:perturbative-probs} into the OT estimator of \cref{eq:povm-estimator}. Using the empirical state frame gives
    \begin{equation}
        \es\mu_a
        = \es\F_\rho^+\es\F_\rho(\mu_a) + \delta\mu_a
        = \mu_a + \delta\mu_a, \qquad
        \delta\mu_a=\frac{1}{\sqrt N}\sum_i\es\xi_{ai}\dualrho_i.
        \label{eq-app:delta-mu}
    \end{equation}
    Conditional independence across training records and \cref{eq-app:conditional-error} then imply
    \begin{equation*}
        \E_\m\!\left[\norm{\delta\Mmu}_F^2 \middle| \rhotr\right]
        = \frac{1}{N} \sum_i\left[1-\sum_a p_a(\rho_i)^2\right]
        \norm{\dualrho_i}_{\mathrm{HS}}^2
        \leq \frac{1}{N}\sum_i\norm{\dualrho_i}_{\mathrm{HS}}^2.
    \end{equation*}
    Finally, $\sum_i\norm{\dualrho_i}_{\mathrm{HS}}^2 =\Tr(\es\F_\rho^+\es\F_\rho\es\F_\rho^+) = \Tr(\es\F_\rho^+)$, which proves the bound.
\end{proof}

In the following, we use the first-order observable error from \cref{prop:ot-linearization} to calculate the MSE to second order in the POVM-estimation error. We first consider OT and then its state-prior variant.

\subsection{Expansion of OT}
\label{app:perturbation-ot}

\Cref{eq-app:delta-mu} shows that OT cancels the noiseless finite-state error exactly: the empirical inverse cancels the frame of the same training sample. Thus $\es\mu_a=\mu_a$ when the training probabilities are exact, already at finite $\ntr$. At finite shots, the same equation also gives $\E_\m[\delta\mu_a\mid\rhotr]=0$.

To compute the remaining contribution, define the rescaled readout frequency error
\begin{equation}
    \es\zeta_i=\bw_\ex^T\es{\bs\xi}_i=\sum_a w_{\ex,a}\es\xi_{ai}.
    \label{eq-app:zeta}
\end{equation}
Substituting \cref{eq-app:delta-mu} into the linearized readout gives
\begin{equation}
    \delta\Ob_{\hw}^{(1)} = -\frac{1}{\sqrt N}\sum_i\es\zeta_i\dualrho_i.
    \label{eq-app:delta-O-ot}
\end{equation}
The ideal weights satisfy $\sum_a w_{\ex,a}\mu_a=\Ob$. Therefore, the full multinomial covariance in \cref{eq-app:conditional-error} gives
\begin{equation}
    \E_\m[\es\zeta_i\mid\rhotr] = 0, \qquad
    \E_\m[\es\zeta_i\es\zeta_j\mid\rhotr] = \delta_{ij}V_\ex(\rho_i,\Ob),
    \label{eq-app:variance-zeta}
\end{equation}
where
\begin{equation*}
    V_\ex(\rho,\Ob) \equiv \bw_\ex^T\bSigma(\bp(\rho))\bw_\ex
    = \sum_a p_a(\rho)(w_{\ex,a})^2-\langle\Ob,\rho\rangle^2
\end{equation*}
is the single-shot variance of the ideal readout~\cite{2007dariano,2014zhu,2023innocentia}, denoted $\Var(\bw_\Ob)$ in Ref.~\cite{2026lomonaco}; it is the reference test variance of Ref.~\cite{2026lomonacoa}.

For fixed training states, the empirical inverse in \cref{eq-app:delta-O-ot} is deterministic. Using \cref{eq-app:variance-zeta}, averaging over the measurement records gives
\begin{equation}
    \MSE_\OT^{(\infty)}(\sigma\mid\rhotr) \simeq\frac{1}{N}\sum_i V_\ex(\rho_i,\Ob) \left\langle\sigma,\dualrho_i\right\rangle^2.
    \label{eq-app:mse-averaging-condition}
\end{equation}
The two factors have a direct interpretation: $V_\ex(\rho_i,\Ob)$ is the measurement noise of the ideal readout, while the squared overlap measures its transfer from the training state to the test state through the dual training frame.

Averaging \cref{eq-app:mse-averaging-condition} over the training and test ensembles, with $\Fte=\E_\sigma[\bv_\sigma\bv_\sigma^T]$, yields
\begin{equation}
    \MSE_\OT^{(\infty)}\simeq\frac{V_\OT}{N\ntr},
    \label{eq-app:mse-ot}
\end{equation}
where, by exchangeability of the training states,
\begin{equation}
    V_\OT = \ntr^2\,\E_\rhotr\!\left[V_\ex(\rho_1,\Ob)\left\langle\dualrho_1,\Fte\!\left(\dualrho_1\right)\right\rangle\right].
    \label{eq-app:VOT-averaged}
\end{equation}
Since $\dualrho_i=O(\ntr^{-1})$, the coefficient $V_\OT$ remains finite as $\ntr$ grows.
The leading error therefore comes from the training-shot fluctuations, whose contribution is determined by the ideal readout variance and the empirical state-frame inverse. There is no independent finite-state contribution that survives for $N\to\infty$, since the empirical inverse cancels the frame of the realized training set.

\subsection{State-prior OT}
\label{app:perturbation-sp}

State-prior OT uses the same empirical POVM readout but replaces the empirical state-frame inverse by the correct population inverse, $\overline{\F}_\rho^{-1}$, as assumed in \cref{eq-app:correct-prior}. Its POVM estimator is $\es\mu_a^\sp=\sum_i\es p_{ai}\overline{\F}_\rho^{-1}(\rho_i)$. Since $\ntr\E_\rho[p_a(\rho)\overline{\F}_\rho^{-1}(\rho)]=\mu_a$, we can decompose its error exactly as
\begin{equation*}
    \es\mu_a^\sp-\mu_a =
    \frac{1}{\sqrt N}\sum_i\es\xi_{ai}\overline{\F}_\rho^{-1}(\rho_i) +
    \frac{1}{\ntr}\sum_i\left[p_a(\rho_i)\,\ntr\overline{\F}_\rho^{-1}(\rho_i)-\mu_a\right] \equiv
    \delta\mu_a^{(N)} + \delta\mu_a^{(\rho)}.
\end{equation*}
The first term $\delta\mu_a^{(N)}$ is the measurement noise, while the second $\delta\mu_a^{(\rho)}$ is the fluctuation of a finite training-state average around its population value, and remains when $N\to\infty$.

Applying \cref{eq-app:delta-oeff} gives the first-order observable error $\delta\Ob_{\hw}^{(1),\sp} = \delta\Ob_N^\sp+\delta\Ob_\rho^\sp$, with
\begin{equation*}
    \delta\Ob_N^\sp = -\frac{1}{\sqrt N} \sum_i\es\zeta_i\overline{\F}_\rho^{-1}(\rho_i), \qquad
    \delta\Ob_\rho^\sp = -\frac{1}{\ntr} \sum_i\mathbb R_\Ob(\rho_i).
\end{equation*}
Here $\es\zeta_i$ is defined in \cref{eq-app:zeta}, and the noiseless state-sampling residual is
\begin{equation}
    \mathbb R_\Ob(\rho) = \langle\Ob,\rho\rangle\,\ntr\overline{\F}_\rho^{-1}(\rho)-\Ob.
    \label{eq-app:state-sampling-residual}
\end{equation}
This follows from $\sum_a w_{\ex,a} p_a(\rho)=\langle\Ob,\rho\rangle$ and $\sum_a w_{\ex,a}\mu_a=\Ob$. The correct-prior assumption gives $\E_\rho[\mathbb R_\Ob(\rho)]=0$.

At fixed training states, $\delta\Ob_\rho^\sp$ is deterministic while $\E_\m[\delta\Ob_N^\sp\mid\rhotr]=0$. The mixed contribution to the quadratic MSE therefore vanishes after averaging over the measurement records. Averaging the two remaining terms gives
\begin{equation}
    \MSE_\sp^{(\infty)}\simeq\frac{V_\sp}{N\ntr}+\frac{V_\rho}{\ntr},
    \label{eq-app:mse-state-prior}
\end{equation}
with
\begin{equation}
    V_\sp =
    \E_\rho\! \left[V_\ex(\rho,\Ob)
    \left\langle\ntr\overline{\F}_\rho^{-1}(\rho),
    \Fte\!\left(\ntr\overline{\F}_\rho^{-1}(\rho)\right)\right\rangle\right], \qquad
    V_\rho =
    \E_\rho\! \left[ \left\langle\mathbb R_\Ob(\rho),
    \Fte(\mathbb R_\Ob(\rho))\right\rangle \right].
    \label{eq-app:state-prior-coefficients}
\end{equation}
Thus, the state prior introduces a sampling contribution that is reduced by increasing $\ntr$, but not by increasing $N$. In particular, increasing the number of shots alone does not ensure that the state-prior estimator lies in the small-perturbation regime.

The shot coefficients of the two methods are not generally equal at finite $\ntr$: \cref{eq-app:VOT-averaged} contains the empirical inverse, whereas \cref{eq-app:state-prior-coefficients} contains the population inverse. In the large-training-set regime, when replacing $\es\F_\rho^+$ by $\overline{\F}_\rho^{-1}$, that is $\dualrho_i$ by $\overline{\F}_\rho^{-1}(\rho_i)$, is justified in these averages, $V_\OT$, which depends on $\ntr$ through the empirical inverse, approaches $V_\sp$. In contrast, $V_\rho$ does not depend on the reconstructing POVM, since the measurement has canceled from \cref{eq-app:state-sampling-residual}.

\subsection{Haar-averaged coefficients and measurement benchmarks}
\label{app:perturbation-benchmarks}

We now evaluate the coefficients that determine the dimension scaling of OT and state-prior OT. We consider Haar-random pure training and test states, writing $\rho_\psi=\ket\psi\!\bra\psi$ and denoting the normalized Haar integral by $\int d\psi$. For OT, we use the population-frame approximation $\es\F_\rho^+\simeq\overline{\F}_\rho^{-1}$ discussed above. Within this approximation, the two methods share the same leading shot-noise coefficient, $V_\OT=V_\sp$, so we will only use the symbols $V_\OT$ and $V_\rho$ for both methods.

We first evaluate the Haar averages for a general measurement. We then calculate the shot coefficient for a balanced 2-design POVM and for the large-outcome limit of the Haar POVMs used in the simulations. The results for the two projector observables are collected in \cref{tab:perturbative-benchmarks}.

\paragraph{Haar averages.}

The test frame coincides with the single-state training frame, $\Fte=\overline{\F}_\rho/\ntr$, with $\overline{\F}_\rho$ given in \cref{eq:haar-state-frame}, so that $\ntr\overline{\F}_\rho^{-1}=\Fte^{-1}$. For a pure state $\rho$, we have
\begin{equation}
    \left\langle\Fte^{-1}(\rho),\Fte\!\left(\Fte^{-1}(\rho)\right)\right\rangle
    = \left\langle\Fte^{-1}(\rho),\rho\right\rangle
    = \left\langle d(d+1)\rho-dI,\rho\right\rangle
    = d^2.
    \label{eq-app:haar-dual-norm}
\end{equation}
Using this identity in \cref{eq-app:VOT-averaged} gives
\begin{equation}
    V_\OT
    = d^2\,\E_\rho[V_\ex(\rho,\Ob)]
    = d^2\int d\psi\left[\sum_a\langle\mu_a,\rho_\psi\rangle(w_{\ex,a})^2-\langle\Ob,\rho_\psi\rangle^2\right].
    \label{eq-app:VOT-haar-integral}
\end{equation}
Thus, the coefficient is $d^2$ times the Haar-averaged single-shot variance of the ideal readout, $V_\OT=d^2V_\ex(\Ob)$ with $V_\ex(\Ob)\equiv\E_\rho[V_\ex(\rho,\Ob)]$, which links it to the reference test variance of Ref.~\cite{2026lomonacoa}. The integral can be evaluated using the first two Haar moments~\cite{2013harrow,2024mele},
\begin{equation}
    \int d\psi\,\langle\mu_a,\rho_\psi\rangle = \frac{\Tr(\mu_a)}{d}, \qquad
    \int d\psi\,\langle\Ob,\rho_\psi\rangle^2 = \frac{(\Tr\Ob)^2+\Tr(\Ob^2)}{d(d+1)}.
    \label{eq-app:haar-moments}
\end{equation}
Substituting into \cref{eq-app:VOT-haar-integral}, we obtain
\begin{equation}
    V_\OT
    = d\sum_a\Tr(\mu_a)(w_{\ex,a})^2-\frac{d}{d+1}\left[(\Tr\Ob)^2+\Tr(\Ob^2)\right].
    \label{eq-app:VOT-empirical-haar}
\end{equation}
This expression applies to any fixed IC POVM. Its dependence on the measurement remains in the effect traces and in the ideal dual weights of \cref{eq-app:ideal-dual-weights}.

\paragraph{Balanced 2-design POVM.}

Consider the measurement $\mu_a=(d/\nout)\Pi_a$, where the projectors $\Pi_a=\ket{\chi_a}\!\bra{\chi_a}$ form a balanced complex projective 2-design~\cite{2004renesa,2006scott,2007roy}. Their first and second moments satisfy
\begin{equation*}
    \frac{1}{\nout}\sum_a\Pi_a = \frac{I}{d}, \qquad
    \frac{1}{\nout}\sum_a\Tr(A\Pi_a)\Pi_a = \frac{\Tr(A)I+A}{d(d+1)}.
\end{equation*}
The measurement frame and its inverse are therefore~\cite{2006scott}
\begin{equation*}
    \F_\mu(A) = \frac{d}{\nout(d+1)}\left[\Tr(A)I+A\right], \qquad
    \F_\mu^{-1}(A) = \frac{\nout}{d}\left[(d+1)A-\Tr(A)I\right].
\end{equation*}
Substituting into \cref{eq-app:ideal-dual-weights} gives $w_{\ex,a}=(d+1)\Tr(\Ob\Pi_a)-\Tr(\Ob)$. Squaring these weights and using the two design identities yields
\begin{equation*}
    \frac{1}{\nout}\sum_a(w_{\ex,a})^2=\frac{(d+1)\Tr(\Ob^2)-(\Tr\Ob)^2}{d}.
\end{equation*}
Since $\Tr(\mu_a)=d/\nout$, the POVM is balanced, so that $V=V_\OT$ and we drop the subscript; \cref{eq-app:VOT-empirical-haar} becomes
\begin{equation}
    V^{\mathrm{2design}}=\frac{d(d+2)}{d+1}\left[d\Tr(\Ob^2)-(\Tr\Ob)^2\right] = \frac{d^3(d+2)}{d+1} \Var(\lambda_\Ob),
    \label{eq-app:2design-VOT}
\end{equation}
where $\Var(\lambda_\Ob)=\Tr(\Ob^2)/d-(\Tr\Ob)^2/d^2$ is the variance of the eigenvalues of $\Ob$.

\paragraph{Haar POVM.}

We now consider the Haar POVM used in \cref{sec:main}. Its effects can be written as $\mu_a = \omega_a\Pi_a$, with $\omega_a\equiv\Tr(\mu_a)$, where each $\Pi_a$ is Haar-distributed and independent of its weight $\omega_a$. Unlike the balanced case, the effect weights are not equal. For a single outcome, their distribution is $\omega_a\sim\operatorname{Beta}(d,\nout-d)$~\cite{2001zyczkowski,2020heinosaari}, which gives
\begin{equation*}
    \E_\mu[\omega_a] = \frac{d}{\nout}, \qquad
    \E_\mu[\omega_a^3] = \frac{d(d+1)(d+2)}{\nout(\nout+1)(\nout+2)}.
\end{equation*}
These are single-outcome moments; the effects of the complete POVM are correlated through $\sum_a\mu_a=I$.

We evaluate the coefficient in the limit of large $\alpha_\mu=\nout/d^2$ at fixed $d$. In this limit the frame concentrates around the ensemble average in \cref{eq:haar-povm-prior}, giving
\begin{equation*}
    \nout\F_\mu(A) \longrightarrow A+\Tr(A)I, \qquad
    \frac{1}{\nout}\F_\mu^{-1}(A) \longrightarrow A-\frac{\Tr(A)}{d+1}I.
\end{equation*}
We use this limit to evaluate the canonical readout coefficient, without changing the training procedure. The ideal weights then have the leading form $w_{\ex,a}\simeq\nout \omega_a[\Tr(\Ob\Pi_a)-\Tr(\Ob)/(d+1)]$. Using exchangeability of the outcomes and independence of each weight from its direction gives
\begin{equation}
    \E_\mu\!\left[\sum_a \omega_a(w_{\ex,a})^2\right]
    \simeq \nout^3\E_\mu[\omega_a^3]
    \int d\psi\left[\langle\Ob,\rho_\psi\rangle-\frac{\Tr(\Ob)}{d+1}\right]^2.
    \label{eq-app:haar-weighted-moment}
\end{equation}
Here the Haar integral is over the measurement direction. Expanding its square and using \cref{eq-app:haar-moments}, we obtain
\begin{equation*}
    \int d\psi\left[\langle\Ob,\rho_\psi\rangle-\frac{\Tr(\Ob)}{d+1}\right]^2
    = \frac{1}{d(d+1)}\left[\Tr(\Ob^2)-\frac{(\Tr\Ob)^2}{d+1}\right].
\end{equation*}
Since $\nout^3\E_\mu[\omega_a^3]\to d(d+1)(d+2)$, \cref{eq-app:haar-weighted-moment} gives
\begin{equation*}
    \E_\mu\!\left[\sum_a \omega_a(w_{\ex,a})^2\right]\longrightarrow(d+2)\left[\Tr(\Ob^2)-\frac{(\Tr\Ob)^2}{d+1}\right].
\end{equation*}
Substituting into \cref{eq-app:VOT-empirical-haar}, we find
\begin{equation}
    V_\OT^{\mathrm{Haar}}
    = \frac{d(d^2+3d+1)}{d+1}\Tr(\Ob^2) - \frac{d(d+3)}{d+1}(\Tr\Ob)^2.
    \label{eq-app:haar-VOT}
\end{equation}
This expression applies in the large-$\alpha_\mu$ limit. It differs from the balanced result of \cref{eq-app:2design-VOT}.

\paragraph{Finite-state coefficient.}

The coefficient $V_\rho$ can be evaluated for the same Haar ensemble without imposing any symmetry on the POVM. Using \cref{eq-app:state-prior-coefficients}, substituting the residual from \cref{eq-app:state-sampling-residual} and using \cref{eq-app:haar-dual-norm} gives
\begin{equation}
    V_\rho
    = (d^2-2)\int d\psi\,\langle\Ob,\rho_\psi\rangle^2+\langle\Ob,\Fte(\Ob)\rangle
    = (d^2-1)\int d\psi\,\langle\Ob,\rho_\psi\rangle^2
    = \frac{d^2-1}{d(d+1)}\left[(\Tr\Ob)^2+\Tr(\Ob^2)\right].
    \label{eq-app:Vrho}
\end{equation}
Here we used $\langle\Ob,\Fte(\Ob)\rangle=\int d\psi\,\langle\Ob,\rho_\psi\rangle^2$ and the second Haar moment. This contribution is independent of the POVM and, as shown in \cref{app:perturbation-ot}, is absent in OT.

\paragraph{Projector observables and dimension scaling.}

As examples we consider the rank-one projector $\proj$ (with unit trace and purity), and the centered projector
\begin{equation}
    \T = \proj - I/d,
    \label{eq-app:center-obs}
\end{equation}
with $\Tr(\T)=0$ and $\Tr(\T^2)=(d-1)/d$.

Substituting these observables into \cref{eq-app:2design-VOT,eq-app:haar-VOT,eq-app:Vrho} gives \cref{tab:perturbative-benchmarks}.

\begin{table*}[t]
    \centering
    \renewcommand{\arraystretch}{1.6}
    \begin{tabular}{lccc}
        \toprule
        Observable & $V^{\mathrm{2design}}$ & $V_\OT^{\mathrm{Haar}}$ & $V_\rho$ \\
        \midrule
        $\proj$ & $\dfrac{d(d+2)(d-1)}{d+1}$ & $\dfrac{d(d^2+2d-2)}{d+1}$ & $\dfrac{2(d-1)}{d}$ \\
        $\T$ & $\dfrac{d(d+2)(d-1)}{d+1}$ & $\dfrac{(d-1)(d^2+3d+1)}{d+1}$ & $\dfrac{(d-1)^2}{d^2}$ \\
        \bottomrule
    \end{tabular}
    \caption{Leading MSE coefficients for Haar-random pure training and test states. The observables considered are the projector $\proj$ and the centered projector $\T = \proj - I/d$ of \cref{eq-app:center-obs}. $V_\OT$ is given for a balanced 2-design POVM and for a Haar POVM at large $\alpha_\mu$. In both cases $V_\OT=O(d^2)$ at large $d$, while $V_\rho=O(1)$.}
    \label{tab:perturbative-benchmarks}
\end{table*}

At fixed $N$ and $\ntr = \alpha_\rho d^2$, \cref{eq-app:mse-ot,eq-app:mse-state-prior} give, for the projector $\proj$ at large $d$,
\begin{equation*}
    \MSE_\OT^{(\infty)} \sim \frac{1}{N\alpha_\rho}, \qquad
    \MSE_\sp^{(\infty)} \sim \frac{1}{N\alpha_\rho}+\frac{2}{\alpha_\rho d^2}.
\end{equation*}
This error does not depend on $d$, but the targets themselves concentrate, $\Var_\rho[\langle\proj,\rho\rangle]=(d-1)/[d^2(d+1)]\sim1/d^2$, so that the error relative to the spread of the targets grows as $d^2$.

\section{Computational complexity of OT}
\label{app:complexity}

In this appendix we analyze the classical computational and memory cost required to construct a single final readout layer, comparing pseudo-inverse training with OT and its prior variants. Throughout the appendix we treat all matrices as dense and represent operators on $\mathcal H$ as vectors in the $d^2$-dimensional operator space.

As a reminder of computational costs~\cite{2012golub}, in dense linear algebra the product of an $r \times s$ matrix and an $s \times t$ matrix requires $O(rst)$ operations. For a dense $r \times s$ matrix, an SVD-based Moore--Penrose inverse has cost $O(rs\min\{r,s\})$. The pseudo-inverse of a square $r \times r$ matrix therefore costs $O(r^3)$. We denote computational costs by $T$ (time) and memory costs by $S$ (space).

The training data are the outcomes of $N$ shots per state, summarized by the empirical probability matrix $\PN \in\RR^{\nout\times\ntr}$ of \cref{eq:empirical-probs}. We do not include the cost of obtaining these frequencies from the raw measurement outcomes, as this is common to every method. The matrix $\PN$ does not need to be stored explicitly in the sequential/online implementations considered below. The training states are collected in the matrix $\Mrho \in\RR^{d^2\times\ntr}$ defined in \cref{eq:probability_matrix}. The single observable is represented by $\bv_\Ob\in\RR^{d^2}$, and the labels are $\by=\Mrho^T\bv_\Ob\in\RR^{\ntr}$. We regard the labels $\by$ as part of the training dataset, so their cost is not included.

Throughout the comparison we consider the regime $\ntr > \nout > d^2$, in which both the training set and the measurement are overcomplete in the operator space.

We first establish the matrix notation and then derive the cost of standard pseudo-inverse training, including its sequential implementation, in \cref{app:complexity-pinv}. We next calculate the corresponding costs for OT and its prior variants in \cref{app:complexity-ot}. Finally, we compare the resource scaling of these methods in \cref{app:complexity-comparison}, identifying when dense OT training is computationally advantageous and discussing the single-shot sparse case.

\subsection{Pseudo-inverse training complexity}
\label{app:complexity-pinv}

Standard pseudo-inverse training, as explained in \cref{sec:qlr-readout}, gives the readout $\hw_\pinv = (\PN^T)^+\by = (\PN^T)^+\Mrho^T\bv_\Ob$. Since $\PN$ is an $\nout\times\ntr$ matrix and computing $\PN^+$ costs $O(\ntr\nout^2)$, with the multiplication the cost of pseudo-inverse training is
\begin{equation}
    T_\pinv = O(\ntr\nout^2) + O(\ntr\nout) = O(\ntr\nout^2).
    \label{eq-app:pinv-cost}
\end{equation}
The main memory requirement comes instead from storing the empirical probability matrix itself, $S_\pinv = O(\ntr\nout)$.

\subsubsection*{Sequential implementation}

Pseudo-inverse training can also be implemented sequentially. Let $\hbp_i$ be the $i$-th column of $\PN$, the empirical probability vector of the training state $\rho_i$. We define the two accumulators
\begin{equation*}
    \widehat{\bs G} = \PN\PN^T = \sum_{i=1}^{\ntr} \hbp_i \hbp_i^T
    \in\RR^{\nout\times\nout}, \qquad
    \widehat{\bs c} = \PN\by = \sum_{i=1}^{\ntr} y_i \hbp_i
    \in\RR^{\nout},
\end{equation*}
which can be updated sequentially with new training states as $\widehat{\bs G} \leftarrow \widehat{\bs G} + \hbp_i\hbp_i^T$ and $\widehat{\bs c} \leftarrow \widehat{\bs c} + y_i\hbp_i$.
Updating $\widehat{\bs G}$ costs $O(\nout^2)$ for every training state, while updating $\widehat{\bs c}$ costs $O(\nout)$. After all $\ntr$ states have been processed, the accumulation therefore costs $O(\ntr\nout^2)$.

Using the identity $(\PN^T)^+=(\PN\PN^T)^+\PN$~\cite{2012barata}, we obtain
\begin{equation*}
    \hw_\pinv = (\PN^T)^+\by = (\PN\PN^T)^+\PN\by = \widehat{\bs G}^+\widehat{\bs c}.
\end{equation*}
Adding the accumulation, the $O(\nout^3)$ pseudo-inverse of $\widehat{\bs G}$, and the final multiplication, we obtain $T_{\pinv, \mathrm{seq}} = O(\ntr\nout^2) + O(\nout^3) = O(\ntr\nout^2)$, which in this regime is the same as \cref{eq-app:pinv-cost}.

The full matrix $\PN$ no longer needs to be stored, only $\widehat{\bs G}$ and $\widehat{\bs c}$, so that $S_{\pinv, \mathrm{seq}} = O(\nout^2)$. Thus, sequential pseudo-inverse training does not change the computational scaling, but it reduces the memory cost.

\subsection{OT training}
\label{app:complexity-ot}

We now analyze the online implementation of OT, summarized in \cref{alg:ot}. OT accumulates the two matrices
\begin{equation}
    \es\F_\rho = \Mrho\Mrho^T = \sum_{i=1}^{\ntr} \bv_{\rho_i}\bv_{\rho_i}^T
    \in\RR^{d^2\times d^2}, \qquad
    \Cm = \PN \Mrho^T = \sum_{i=1}^{\ntr} \hbp_i\bv_{\rho_i}^T
    \in\RR^{\nout\times d^2}.
    \label{eq-app:accumulators-ot}
\end{equation}
The first accumulator is the empirical state frame of \cref{eq:state-frame} in vectorized form, while the second collects the training states weighted by the observed frequencies. Both are sums over the training states, so they can be updated as new events arrive without storing $\PN$ or $\Mrho$, and the readout can be computed from them at any time, as in \cref{alg:ot}.

\begin{algorithm}[H]
    \centering
    \begin{minipage}{0.5\textwidth}
    \fontsize{10pt}{15pt}\selectfont\setlength{\lineskiplimit}{-\maxdimen}
    \begin{algorithmic}[1]
        \Require states $\rho_i$, counts $\hbn_i$, shots $N$, target $\bv_\Ob$
        \Ensure readout $\hw_\OT$
        \State $\es\F_\rho \leftarrow 0$, \quad $\Cm \leftarrow 0$
        \For{$i=1,\ldots,\ntr$}
            \State $\es\F_\rho \leftarrow \es\F_\rho+\bv_{\rho_i}\bv_{\rho_i}^T$
            \State $\Cm \leftarrow \Cm+\hbn_i\bv_{\rho_i}^T/N$
        \EndFor
        \State $\hatMmu^T \leftarrow \Cm \es\F_\rho^+$
        \State \Return $\hw_\OT = \hatMmu^+\bv_\Ob$
    \end{algorithmic}
    \end{minipage}
    \caption{Online Training (OT). The accumulators $\es\F_\rho$ and $\Cm$ of \cref{eq-app:accumulators-ot} are updated with each training state and its counts, without storing $\PN$ or $\Mrho$, and the readout can be computed from them at any time. The total cost is $O(\ntr\nout d^2)$ and the memory $O(d^4)+O(\nout d^2)$.}
    \label{alg:ot}
\end{algorithm}

We collect the estimated POVM elements as columns of $\hatMmu$, introduced in \cref{eq:explicit_estimator_OT}, which can be written as
\begin{equation*}
    \hatMmu^T = \PN \Mrho^+
    = \PN \Mrho^T (\Mrho\Mrho^T)^+
    = \Cm \es\F_\rho^+,
\end{equation*}
where we used the identity $\Mrho^+=\Mrho^T(\Mrho\Mrho^T)^+$~\cite{2012barata}. This is the matrix form of the POVM estimator in \cref{eq:povm-estimator}.

The estimated POVM frame is $\es\F_\mu = \hatMmu\hatMmu^T$, so that the canonical dual readout is $\hw_\OT = \hatMmu^T \es \F_\mu^+ \bv_\Ob = \hatMmu^T (\hatMmu \hatMmu^T)^+ \bv_\Ob$, which equals the compact expression $\hatMmu^+\bv_\Ob$ of \cref{eq:explicit_estimator_OT}.

We now count the operations of the online algorithm. Each update of $\es\F_\rho$ in \cref{eq-app:accumulators-ot} costs $O(d^4)$ and each update of $\Cm$ costs $O(\nout d^2)$, so that processing all the $\ntr$ states costs $O(\ntr\nout d^2) + O(\ntr d^4)$. The pseudo-inverse $\es\F_\rho^+$ costs $O(d^6)$, the product with $\Cm$ costs $O(\nout d^4)$, and the pseudo-inverse of the estimated $\hatMmu^T\in\RR^{\nout\times d^2}$ costs another $O(\nout d^4)$. In total, in this regime,
\begin{equation*}
    T_\OT = O(\ntr\nout d^2) + O(\ntr d^4) + O(\nout d^4) + O(d^6) = O(\ntr\nout d^2).
\end{equation*}
In memory, only $\es\F_\rho$ and $\Cm$ are kept, so that $S_\OT = O(d^4) + O(\nout d^2)$.

\subsubsection*{Savings from prior methods}

The priors of \cref{sec:priors} replace the empirical frames with their population averages.

With state-prior OT, the estimated POVM is $(\hatMmu^\sp)^T = \Cm \overline{\F}_\rho^{-1}$. For a generic prior this product would cost $O(\nout d^4)$, but the Haar prior of \cref{eq:haar-state-frame} acts as $\overline{\F}_\rho^{-1}(A)=[d(d+1)A-d\Tr(A)I]/\ntr$, so that applying it to each of the $\nout$ rows of $\Cm$ costs only $O(d^2)$, for a total of $O(\nout d^2)$. This also removes the $O(\ntr d^4)$ accumulation of $\es\F_\rho$ and its $O(d^6)$ inversion. All the rest is the same as OT, so that $T_\sp = O(\ntr\nout d^2) + O(\nout d^4)$ and $S_\sp = O(\nout d^2)$.

POVM-prior OT avoids the empirical dual inversion but still reconstructs $\hatMmu$ with $\es\F_\rho^+$, so that $T_\mup = T_\OT$ and $S_\mup = S_\OT$. Double-prior OT has both savings, so that $T_\text{d-p} = O(\ntr\nout d^2)$ and $S_\text{d-p} = O(\nout d^2)$.

\subsection{Computational advantage of OT}
\label{app:complexity-comparison}

\begin{table*}[t]
    \centering
    \small
    \begin{tabular}{lcccc}
        \toprule
        Method
        & Accumulators
        & Inversions
        & Computational cost
        & $\alpha_\rho \gg \alpha_\mu$\\
        \midrule
        Pseudo-inverse
        & $\widehat{\bs G},\ \widehat{\bs c}$
        & $\widehat{\bs G}\in\RR^{\nout\times\nout}$
        & $O(\ntr\nout^2+\nout^3)$
        & $O(\alpha_\rho\alpha_\mu^2 d^6)$ \\
        OT
        & $\es\F_\rho,\ \Cm$
        & $\es\F_\rho,\ \hatMmu$
        & $O(\ntr\nout d^2+\ntr d^4+\nout d^4+d^6)$
        & $O(\alpha_\rho\alpha_\mu d^6)$ \\
        State-prior OT
        & $\Cm$
        & $\hatMmu^\sp$
        & $O(\ntr\nout d^2+\nout d^4)$
        & $O(\alpha_\rho\alpha_\mu d^6)$ \\
        POVM-prior OT
        & $\es\F_\rho,\ \Cm$
        & $\es\F_\rho$
        & $O(\ntr\nout d^2+\ntr d^4+\nout d^4+d^6)$
        & $O(\alpha_\rho\alpha_\mu d^6)$ \\
        Double-prior OT
        & $\Cm$
        & none
        & $O(\ntr\nout d^2)$
        & $O(\alpha_\rho\alpha_\mu d^6)$ \\
        \bottomrule
    \end{tabular}
    \caption{Computational cost of constructing a single final readout. The second column shows the accumulators stored in memory, and the third the matrices that are pseudo-inverted. The fourth column gives the total cost for $\ntr>\nout>d^2$ and dense empirical probability vectors, and the last column its leading form for $\alpha_\rho = \ntr/d^2 \gg \alpha_\mu = \nout/d^2$.}
    \label{tab:complexity-summary}
\end{table*}

Collecting the results, the dense costs for constructing a single final readout are summarized in \cref{tab:complexity-summary}. To compare pseudo-inverse training and OT, we use the overcompleteness parameters $\alpha_\rho=\ntr/d^2$ and $\alpha_\mu=\nout/d^2$ of \cref{app:priors}, which compare the training set and the effective POVM with an IC frame of $d^2$ elements~\cite{2023innocentia}. In the regime $\alpha_\rho \gg \alpha_\mu \gg 1$ of many more training states than outcomes, shown in the last column of \cref{tab:complexity-summary},
\begin{equation*}
    T_\pinv = O(\alpha_\rho\alpha_\mu^2d^6), \qquad
    T_\OT = O(\alpha_\rho\alpha_\mu d^6).
\end{equation*}
This shows the computational advantage of OT by a factor of $\alpha_\mu = \nout / d^2$. The origin of this difference is visible from the sequential updates. While pseudo-inverse training accumulates $\hbp_i\hbp_i^T$, which costs $O(\nout^2)$ for every training state, OT accumulates $\hbp_i\bv_{\rho_i}^T$, which costs $O(\nout d^2)$.

The memory cost shows a similar advantage, $S_{\pinv, \mathrm{seq}} = O(\alpha_\mu^2 d^4)$ against $S_\OT = O((1+\alpha_\mu)d^4)$.

Finally, all these calculations assume dense matrices. Sparse matrices can further reduce the cost. For example, in the single-shot regime the columns of $\PN$ are one-hot vectors and $\widehat{\bs G}$ is diagonal, and one finds $T_\pinv^{N=1} = O(\alpha_\rho d^2)$ and $T_\OT^{N=1} = O(\alpha_\rho d^6)$. Thus, pseudo-inverse training can be computationally cheap in the single-shot regime, although this is where its error saturates at the plateau shown in \cref{fig:mse-simulations}.

\section{MSE decomposition for double-prior OT}
\label{app:analytical}

In this appendix we derive the decomposition introduced in \cref{eq:bias-coefficients} for the error $\MSE^{(\infty)}$ with exact test probabilities defined in \cref{eq:mse-inf}. The quantity in \cref{eq:mse-inf} is the error of a fixed trained readout, while here we additionally average over the training realizations, i.e.\ over both the sampled training states and their measurement outcomes. We consider first the estimator where the state-frame and POVM-frame inverses are fixed to the ensemble priors $\overline{\F}_\rho^{-1}$ and $\overline{\F}_\mu^{-1}$ introduced in \cref{sec:priors}. The empirical-frame variants are treated separately in \cref{app:perturbation}.

We first express the prediction as an average of independent training contributions. We then identify the asymptotic bias, the conditional shot-noise variance, and the finite-state sampling variance in \cref{app:analytical-coefficients}. Finally, we combine these terms and average over the test-state ensemble in \cref{app:analytical-mse} to obtain the full MSE decomposition.

We consider a fixed effective POVM $\{\mu_a\}_{a=1}^{\nout}$ and a target observable $\Ob$. Following \cref{sec:qlr-readout}, we write $p_a(\rho) = \langle \mu_a,\rho\rangle$ for the probability of outcome $a$ on a state $\rho$ and $\F_\mu(A) = \sum_a\langle\mu_a,A\rangle\mu_a$ for the frame of the true POVM. The training states $\rho_i$ are sampled independently from the training ensemble, and each state is measured $N$ times. The corresponding counts $\es n_{ai}$ are multinomially distributed and satisfy $\E_{\m}\!\left[\es n_{ai}\mid\rho_i\right] = Np_a(\rho_i)$.

We also assume that the training-state prior $\overline{\F}_\rho$ is the correct one, namely
\begin{equation}
    \overline{\F}_\rho = \E[\es\F_\rho] = \ntr\,\E_\rho [ \bv_\rho\bv_\rho^T],
    \label{eq-app:correct-prior}
\end{equation}
which, for the global Haar ensemble, is the frame of \cref{eq:haar-state-frame}.

With the fixed state-prior inverse, the POVM estimator from \cref{eq:povm-estimator} is
\begin{equation}
    \es \mu_a^\sp = \frac{1}{N} \sum_{i=1}^{\ntr}  \es n_{ai}\, \overline{\F}_\rho^{-1}(\rho_i),
    \label{eq-app:povm-estimator}
\end{equation}
and the readout is then constructed using the prior POVM-frame inverse as $\es w_a = \langle\es\mu_a^\sp,\overline{\F}_\mu^{-1}(\Ob)\rangle$, the empirical counterpart of the prior weights $\overline{w}_a = \langle\mu_a,\overline{\F}_\mu^{-1}(\Ob)\rangle$.

For a test state $\sigma$, the prediction with exact test probabilities $\bp(\sigma)$ is $\es y(\sigma) = \hw^T\bp(\sigma) = \sum_a p_a(\sigma)\es w_a$. Substituting \cref{eq-app:povm-estimator}, we obtain
\begin{align*}
    \es y(\sigma)
    &= \sum_a p_a(\sigma)
    \left\langle
        \frac{1}{N}\sum_{i=1}^{\ntr}\es n_{ai}\overline{\F}_\rho^{-1}(\rho_i),
        \overline{\F}_\mu^{-1}(\Ob)
    \right\rangle \nonumber\\
    &= \frac{1}{\ntr}\sum_{i=1}^{\ntr}
    \left\langle
        \ntr\overline{\F}_\rho^{-1}(\rho_i),
        \overline{\F}_\mu^{-1}(\Ob)
    \right\rangle
    \left(
        \frac{1}{N}\sum_a \es n_{ai}p_a(\sigma)
    \right).
\end{align*}
It is therefore useful to define
\begin{equation*}
    A_i \equiv \left\langle\ntr\overline{\F}_\rho^{-1}(\rho_i),\overline{\F}_\mu^{-1}(\Ob)\right\rangle, \qquad
    B_i(\sigma) \equiv \hbp_i^T\bp(\sigma) = \frac{1}{N}\sum_a \es n_{ai}p_a(\sigma).
\end{equation*}
With this notation,
\begin{equation}
    \es y(\sigma) = \frac1\ntr \sum_{i=1}^{\ntr} A_i B_i(\sigma).
    \label{eq-app:y-average}
\end{equation}
The two factors separate the two sources of randomness that will enter the variance. The quantity $A_i$, which does not depend on $\ntr$ since $\ntr\overline{\F}_\rho^{-1}$ is the inverse of the single-state frame, depends only on the sampled training state $\rho_i$, while $B_i(\sigma)$ also fluctuates through the finite measurement outcomes. Since the training states and their measurements are independent across $i$, the terms $A_iB_i(\sigma)$ are independent and identically distributed.

\subsection{The three error coefficients}
\label{app:analytical-coefficients}

\subsubsection{\texorpdfstring{$\beta^2$}{beta squared}, mean prediction and asymptotic bias}

We start by calculating the conditional mean for a fixed training state $\rho_i=\rho$. In this case,
\begin{equation*}
    A(\rho) \equiv \left. A_i\right|_{\rho_i=\rho}
    = \left\langle\ntr\overline{\F}_\rho^{-1}(\rho),\overline{\F}_\mu^{-1}(\Ob)\right\rangle, \qquad
    B(\rho,\sigma) \equiv \E_{\hbn_i}\!\left[B_i(\sigma)\mid\rho_i=\rho\right]
    = \bp(\rho)^T\bp(\sigma).
\end{equation*}
Using the definition of the POVM frame, the second quantity can also be written as $B(\rho,\sigma) = \langle\rho,\F_\mu(\sigma)\rangle$.

We can now average over the training data $\mathcal D_\mathrm{tr}$ of \cref{eq:mse-avg}. Since all terms in \cref{eq-app:y-average} have the same expectation,
\begin{equation*}
    \overline y(\sigma)
    \equiv \E_{\mathcal D_\mathrm{tr}}[\es y(\sigma)] = \E_\rho[A(\rho)B(\rho,\sigma)]
    = \E_\rho \left[
        \left\langle\ntr\overline{\F}_\rho^{-1}(\rho),\overline{\F}_\mu^{-1}(\Ob)\right\rangle
        \left\langle\rho,\F_\mu(\sigma)\right\rangle
    \right].
\end{equation*}
Using the self-adjointness of the frame operators together with the correct-prior assumption of \cref{eq-app:correct-prior}, the state average cancels the state-prior inverse and gives
\begin{equation*}
    \overline y(\sigma)
    = \left\langle\overline{\F}_\mu^{-1}(\Ob),\F_\mu(\sigma)\right\rangle
    = \left\langle\sigma,\F_\mu\overline{\F}_\mu^{-1}(\Ob)\right\rangle.
\end{equation*}

Therefore, even after averaging over infinitely many training states and measurement outcomes, the prior readout does not in general reconstruct $\Ob$. Instead, it reconstructs the effective observable $\Ob_{\overline{\bw}} = \sum_a\overline{w}_a\mu_a = \F_\mu\overline{\F}_\mu^{-1}(\Ob)$, which differs from the target by $\Delta\bar\Ob \equiv \Ob-\Ob_{\overline{\bw}} = (I-\F_\mu\overline{\F}_\mu^{-1})(\Ob)$.

This is the origin of an asymptotic bias that we will call
\begin{equation*}
    \beta^2(\sigma) \equiv \langle\sigma,\Delta\bar\Ob\rangle^2
    = \left\langle\sigma,\left(I-\F_\mu\overline{\F}_\mu^{-1}\right)(\Ob)\right\rangle^2.
\end{equation*}
For a test ensemble with frame $\Fte = \E_\sigma [\bv_\sigma\bv_\sigma^T]$, we can average this contribution over the test states and obtain
\begin{equation}
    \beta^2 = \left\langle\Delta\bar\Ob,\Fte(\Delta\bar\Ob)\right\rangle.
    \label{eq-app:bias}
\end{equation}
This term has a different nature from the other two contributions we will see later. It does not originate from statistical fluctuations of the training data and is therefore independent of both $N$ and $\ntr$. It measures the prediction error caused by using the ensemble-average POVM frame $\overline{\F}_\mu$ for a specific finite measurement with frame $\F_\mu$.

The observable mismatch contributes to the MSE only when it changes the predictions on the test ensemble. In particular, $\beta^2=0$ precisely when $\langle\sigma,\Ob_{\overline{\bw}}\rangle=\langle\sigma,\Ob\rangle$ for almost every test state. Otherwise, $\beta^2>0$ and the MSE approaches this nonzero value as the resources ($N$ and $\ntr$) are increased.

\subsubsection{\texorpdfstring{$V_\OT$}{V\_OT}, conditional shot-noise variance}

We now consider the fluctuations due to the finite number of measurements performed on each training state. For fixed $\rho_i=\rho$, the factor $A_i=A(\rho)$ is deterministic, while $B_i(\sigma)$ is the average of $N$ outcomes taking the value $p_a(\sigma)$ with probability $p_a(\rho)$. Its conditional variance is
\begin{equation*}
    \Var_{\hbn_i}\!\left(B_i(\sigma)\mid\rho_i=\rho\right)
    = \frac{1}{N}\left[\sum_a p_a(\rho)p_a(\sigma)^2-B(\rho,\sigma)^2\right].
\end{equation*}
Since $A_i=A(\rho)$ is fixed under the same conditioning, the conditional variance of $A_iB_i(\sigma)$ is $A(\rho)^2$ times this one. Averaging it over the training-state ensemble gives the first contribution in the law of total variance,
\begin{equation}
    \E_\rho [\Var_{\hbn_i}(A_iB_i(\sigma)\mid\rho_i=\rho)] = \frac{V_\OT(\sigma)}{N}, \qquad
    V_\OT(\sigma) \equiv \E_\rho
    \left[ A(\rho)^2 \left( \sum_a p_a(\rho) p_a(\sigma)^2 - B(\rho,\sigma)^2 \right) \right].
    \label{eq-app:total-variance-VOT}
\end{equation}
The coefficient $V_\OT(\sigma)$ therefore measures the average single-training-state shot variance. The factor $1/N$ comes from averaging the $N$ measurements performed on each state. Since the final estimator in \cref{eq-app:y-average} averages over $\ntr$ independent training states, this contribution will appear in the MSE as $V_\OT/(N\ntr)$.

\subsubsection{\texorpdfstring{$V_\rho$}{V\_rho}, finite training-state sampling variance}

The second contribution to the variance comes from the fact that the training states themselves form a finite random sample. For a fixed state $\rho_i=\rho$, the conditional mean of a single training contribution is $\E_{\hbn_i}[A_iB_i(\sigma)\mid\rho_i=\rho] = A(\rho)B(\rho,\sigma)$. Its fluctuations over the training-state ensemble are therefore quantified by
\begin{equation}
    V_\rho(\sigma) \equiv \Var_\rho\!\left(A(\rho)B(\rho,\sigma)\right)
    = \E_\rho\left[
        \left( A(\rho)B(\rho,\sigma) - \overline y(\sigma) \right)^2
    \right].
    \label{eq-app:Vrho-sigma}
\end{equation}
Unlike $V_\OT$, this term is already present when $N\to\infty$. It is the sampling variance produced by replacing the population average over the training-state ensemble with an average over a finite set of $\ntr$ states. Since \cref{eq-app:y-average} averages $\ntr$ independent contributions, the corresponding MSE term scales as $V_\rho/\ntr$.

This finite-state contribution is a consequence of using the fixed inverse $\overline{\F}_\rho^{-1}$. OT instead constructs the empirical state-frame inverse $\es\F_\rho^+$ from the same realization of the training states. As shown perturbatively in \cref{app:perturbation}, this adaptation cancels the leading finite-state fluctuation responsible for the $V_\rho$ term.

\subsection{Decomposition of the MSE}
\label{app:analytical-mse}

The mean squared error on a test state $\sigma$ with exact test probabilities is $\MSE^{(\infty)}(\sigma) = \E_{\mathcal D_\mathrm{tr}} [(\es y(\sigma)-y(\sigma))^2]$, where $\mathcal D_\mathrm{tr}=\{(\rho_i,\hbn_i)\}_{i=1}^{\ntr}$ are the training data, as in \cref{eq:mse-avg}. Using the usual mean--variance decomposition,
\begin{equation*}
    \MSE^{(\infty)}(\sigma) = \left(\E_{\mathcal D_\mathrm{tr}}[\es y(\sigma)]-y(\sigma)\right)^2
    + \Var_{\mathcal D_\mathrm{tr}}(\es y(\sigma)).
\end{equation*}
The first term is precisely $\beta^2(\sigma)$. For the second one, independence of the $\ntr$ training contributions in \cref{eq-app:y-average} gives $\Var_{\mathcal D_\mathrm{tr}}(\es y(\sigma)) = \Var_{\mathcal D_\mathrm{tr}}(A_iB_i(\sigma))/\ntr$. We then apply the law of total variance to a single training contribution,
\begin{equation*}
    \Var_{\mathcal D_\mathrm{tr}}\!\left(A_iB_i(\sigma)\right)
    = \E_\rho\left[ \Var_{\hbn_i}\!\left(A_iB_i(\sigma)\mid\rho_i=\rho\right) \right]
    + \Var_\rho\left( \E_{\hbn_i}\!\left[A_iB_i(\sigma)\mid\rho_i=\rho\right] \right)
    = \frac{V_\OT(\sigma)}{N}+V_\rho(\sigma),
\end{equation*}
where we used \cref{eq-app:total-variance-VOT,eq-app:Vrho-sigma}. We therefore obtain
\begin{equation*}
    \MSE^{(\infty)}(\sigma) = \beta^2(\sigma) + \frac{V_\OT(\sigma)}{N\ntr} + \frac{V_\rho(\sigma)}{\ntr}.
\end{equation*}

Finally, averaging over the test-state ensemble gives
\begin{equation}
    \MSE^{(\infty)}(N,\ntr) = \E_\sigma\!\left[\MSE^{(\infty)}(\sigma)\right] = \beta^2+\frac{V_\OT}{N\ntr}+\frac{V_\rho}{\ntr},
    \label{eq-app:MSE}
\end{equation}
where $\beta^2=\E_\sigma[\beta^2(\sigma)]$, $V_\OT=\E_\sigma[V_\OT(\sigma)]$ and $V_\rho=\E_\sigma[V_\rho(\sigma)]$, with $\beta^2$ equivalently given by \cref{eq-app:bias}. This is the training-averaged error of \cref{eq:mse-avg}: it includes both the training average and the average over test states, while excluding finite-shot noise at inference.

The three terms have a direct statistical interpretation. The term $\beta^2$ is the systematic error produced by replacing the frame of the actual finite POVM with its ensemble-average prior, and therefore remains when both $N$ and $\ntr$ are taken to infinity. The term $V_\OT/(N\ntr)$ comes from the multinomial fluctuations of the training measurements and can be reduced by increasing either the number of shots per state or the number of training states. The term $V_\rho/\ntr$ instead comes from the finite sampling of the training-state ensemble and remains even for exact training probabilities.

This calculation uses fixed frame inverses. For an IC true POVM, learning the POVM-frame inverse $\es\F_\mu^+$ removes the prior asymptotic bias $\beta^2$ as the measurement estimate converges. State-prior OT still has the finite-state contribution $V_\rho$, while OT cancels it by using the frame of the realized spanning training set. The corresponding leading MSEs of OT and state-prior OT are derived perturbatively in \cref{app:perturbation}.

\section{Validity of the prior methods}
\label{app:priors}

In this appendix we study when the prior versions of OT approximate OT. Replacing an empirical frame inverse by an analytic prior reduces the computational cost, as discussed in \cref{app:complexity}, but introduces an additional statistical error. The goal of this appendix is to show that this additional cost is controlled and has a simple dependence on the experimental resources.

The two replacements are summarized in \cref{fig:ot-variants}. State-prior OT uses $\overline{\F}_\rho^{-1}$ instead of $\es\F_\rho^+$, while POVM-prior OT uses the inverse of the ensemble-average POVM frame, $\overline{\F}_\mu^{-1}$, instead of $\es\F_\mu^+$. Their controlling resources are the dimension-normalized sample sizes $\alpha_\rho = \ntr/d^2$ and $\alpha_\mu = \nout/d^2$, which measure the overcompleteness of the training-state and POVM frames, respectively. We consider $\alpha_\rho,\alpha_\mu\geq1$ and retain the abbreviations $\sp$ and $\mup$ for the two prior methods.

The leading excess errors behave as $\Delta_\sp\propto\alpha_\rho^{-1}$ and $\Delta_\mup\propto\alpha_\mu^{-1}$. The first is the finite-state contribution derived in \cref{app:perturbation-sp}. The second is the systematic error caused by replacing the frame of the realized POVM by its ensemble average, identified by $\beta^2$ in \cref{app:analytical}. As in \cref{eq-app:MSE}, the analytical MSEs are averaged over training realizations and test states. Additional averages over the random POVM and observable direction will be written explicitly below.

We first study how the finite state and POVM frames approach their ensemble averages in \cref{app:priors-concentration}. We then examine the excess MSE introduced by the state prior as the training set grows in \cref{app:priors-state-excess}. Finally, we analyze the POVM-prior excess MSE as the measurement overcompleteness increases in \cref{app:priors-povm-excess}. In both error studies, we compare the analytical predictions with numerical results.

\subsection{Concentration of state and POVM frames}
\label{app:priors-concentration}

\begin{figure*}
    \centering
    \includegraphics[width=0.75\textwidth]{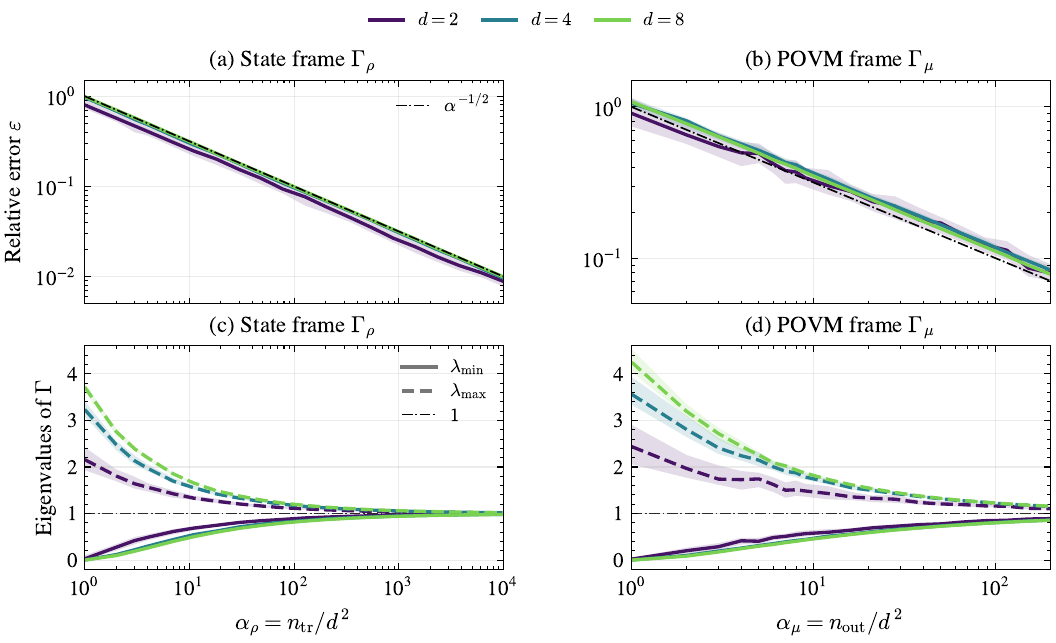}
    \caption{Concentration of the state frame, (a) and (c), and of the POVM frame, (b) and (d), for $d=2,4,8$. (a), (b) Relative Frobenius distances $\varepsilon$ of $\Gamma_\rho$ and $\Gamma_\mu$ from the identity; the dash-dotted lines show the scaling $\alpha^{-1/2}$ of \cref{eq-app:frame-concentration}. (c), (d) Smallest (solid) and largest (dashed) eigenvalues of $\Gamma_\rho$ and $\Gamma_\mu$, which approach one (dash-dotted line) as $\alpha_\rho$ and $\alpha_\mu$ increase. The POVM frame is evaluated directly from the generated measurement effects, without training-shot noise. Curves show the median over 100 independent repetitions and shaded bands the 30th to 70th percentiles.}
    \label{fig:frame-distances}
\end{figure*}

For the state prior, we compare the frame of a finite sample of Haar-random training states with its population frame~\cite{2005ledoux,2011vershynin,2012tropp,2015tropp}. For the POVM prior, we compare the frame of a finite Haar POVM realization with its ensemble average~\cite{2020heinosaari}. The relative frames are
\begin{equation*}
    \Gamma_\rho = \overline{\F}_\rho^{-1/2} \es\F_\rho \overline{\F}_\rho^{-1/2}, \qquad
    \Gamma_\mu = \overline{\F}_\mu^{-1/2} \F_\mu \overline{\F}_\mu^{-1/2},
\end{equation*}
which both satisfy $\E[\Gamma]=I$ since the population frames are the ensemble averages of the realized frames $\es\F_\rho$ and $\F_\mu$.
Then we measure the relative Frobenius distance
\begin{equation*}
    \varepsilon_\rho = \frac{\norm{\Gamma_\rho-I}_F}{\norm{I}_F}
    = \frac1d \sqrt{\sum_{j=1}^{d^2}(\lambda^\rho_j - 1)^2} \,, \qquad
    \varepsilon_\mu = \frac{\norm{\Gamma_\mu-I}_F}{\norm{I}_F}
    = \frac1d \sqrt{\sum_{j=1}^{d^2}(\lambda^\mu_j - 1)^2} \,,
\end{equation*}
where $\{\lambda_j^\rho\}_{j=1}^{d^2}$ and $\{\lambda_j^\mu\}_{j=1}^{d^2}$ are the eigenvalues of $\Gamma_\rho$ and $\Gamma_\mu$, respectively. The identity in these expressions acts on the $d^2$-dimensional operator space, so $\norm{I}_F=d$. The numerical results in \cref{fig:frame-distances} are consistent with the asymptotic scalings
\begin{equation}
    \varepsilon_\rho \sim \alpha_\rho^{-1/2}, \qquad \varepsilon_\mu \sim \alpha_\mu^{-1/2}.
    \label{eq-app:frame-concentration}
\end{equation}
\Cref{tab:frame-distances} reports the fitted exponents and amplitudes. In the perturbative regime, the reconstructed-observable error is linear in the leading frame fluctuation~\cite{1973wedin,2019xu}. Its squared contribution to the MSE therefore scales with the inverse sample size, as examined in the following subsections.

\begin{table}
    \centering
    \small
    \begin{tabular}{l@{\hskip 1em}c@{\hskip 1em}cc}
        \toprule
        Frame & $d$ & Exponent $s$ & Prefactor $C$ \\
        \midrule
        State frame $\Gamma_\rho$ & $2$ & $-0.496$ & $0.815$ \\
        State frame $\Gamma_\rho$ & $4$ & $-0.500$ & $0.967$ \\
        State frame $\Gamma_\rho$ & $8$ & $-0.500$ & $0.993$ \\
        \midrule
        POVM frame $\Gamma_\mu$ & $2$ & $-0.455$ & $0.940$ \\
        POVM frame $\Gamma_\mu$ & $4$ & $-0.487$ & $1.112$ \\
        POVM frame $\Gamma_\mu$ & $8$ & $-0.497$ & $1.096$ \\
        \bottomrule
    \end{tabular}
    \caption{Frame-concentration fits to $\varepsilon=Cx^s$, with $x=\alpha_\rho$ for the state frame and $x=\alpha_\mu$ for the POVM frame. The asymptotic exponent in \cref{eq-app:frame-concentration} is $-1/2$.}
    \label{tab:frame-distances}
\end{table}

\begin{figure*}
    \centering
    \includegraphics[width=0.75\textwidth]{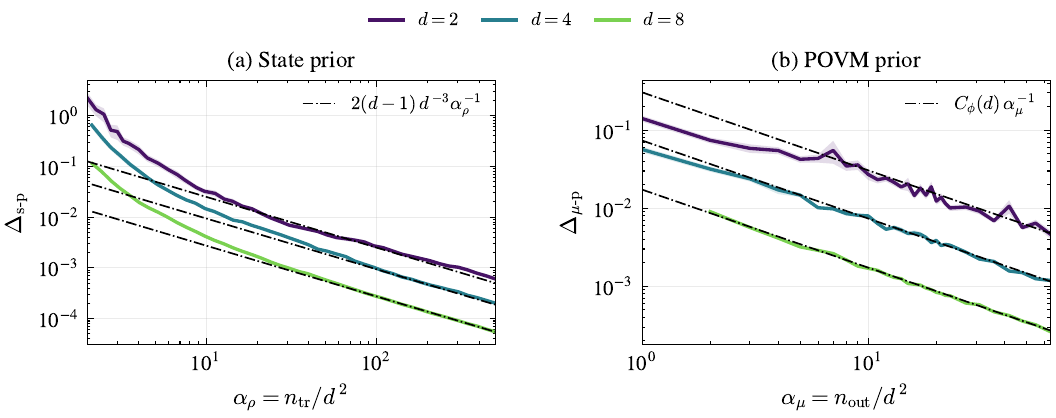}
    \caption{Prior-approximation excess MSE for the rank-one projector $\proj$, for $d=2,4,8$.
    (a) State-prior excess
    $\Delta_\sp=\MSE_\sp^{(\infty)}-\MSE_\OT^{(\infty)}$ against $\alpha_\rho=\ntr/d^2\geq2$,
    using $N=10\,000$ shots per training state and $\nout=4d^2$.
    (b) POVM-prior excess
    $\Delta_\mup=\MSE_\mup^{(\infty)}-\MSE_\OT^{(\infty)}$ against $\alpha_\mu=\nout/d^2$,
    using $N=1\,000$ shots per training state and $\ntr=500d^2$.
    Dash-dotted lines show the analytical predictions of
    \cref{eq-app:state-prior-mse,eq-app:povm-prior-excess-phi}.
    Curves show the mean over 100 independent repetitions, and shaded bands indicate one standard error of the mean.}
    \label{fig:prior-excess}
\end{figure*}

\subsection{State-prior excess MSE}
\label{app:priors-state-excess}

We now study the MSE cost of replacing the empirical state-frame inverse by the analytic prior, through the excess error $\Delta_\sp(\alpha_\rho,d) \equiv \MSE_\sp^{(\infty)}(\alpha_\rho,d) - \MSE_\OT^{(\infty)}(\alpha_\rho,d)$. This difference measures the effect of replacing $\es\F_\rho^+$ by $\overline{\F}_\rho^{-1}$. Subtracting the perturbative expressions in \cref{eq-app:mse-state-prior,eq-app:mse-ot} gives $\Delta_\sp(\alpha_\rho,d) \simeq V_\rho/(\alpha_\rho d^2)$.

For Haar-random pure training and test states, \cref{eq-app:Vrho} gives the finite-state coefficient without requiring a balanced measurement. It therefore applies to the random Haar POVMs used here. For the rank-one projector $\proj$ and its traceless component $\T$ of \cref{eq-app:center-obs}, we obtain
\begin{equation}
    \Delta_{\sp,\,\phi} \simeq \frac{2(d-1)}{d^3}\frac{1}{\alpha_\rho},
    \qquad
    \Delta_{\sp,\,\T} \simeq \frac{(d-1)^2}{d^4}\frac{1}{\alpha_\rho}.
    \label{eq-app:state-prior-mse}
\end{equation}
\Cref{fig:prior-excess}(a) compares the projector data with the first expression and shows good agreement with the predicted inverse-$\alpha_\rho$ behavior. The corresponding fits are summarized in \cref{tab:mse-excess}.

\subsection{POVM-prior excess MSE}
\label{app:priors-povm-excess}

We now study the cost of using the prior POVM frame, through the excess error $\Delta_\mup(\alpha_\mu,d) \equiv \MSE_\mup^{(\infty)}(\alpha_\mu,d) - \MSE_\OT^{(\infty)}(\alpha_\mu,d)$. In the limit of accurate measurement reconstruction, this difference approaches the systematic contribution $\beta^2$ of \cref{app:analytical}. At finite training resources it can also contain shot-dependent corrections. The large values of $N$ and $\ntr$ used here are chosen to make the prior mismatch the leading contribution.

Since here the observable varies while the test ensemble is fixed, we write the observable as a subscript, $\beta^2_\Ob$. For a fixed POVM realization, \cref{eq-app:bias} with the Haar test frame $\Fte=\overline{\F}_\rho/\ntr$ of \cref{eq:haar-state-frame} gives
\begin{equation*}
    \beta^2_\Ob = \frac{\Tr(\Delta\bar\Ob^2)+(\Tr\Delta\bar\Ob)^2}{d(d+1)}.
\end{equation*}
With the prior weights of \cref{app:analytical}, now written as $\overline{w}_a(\Ob)=\langle\mu_a,\overline{\F}_\mu^{-1}(\Ob)\rangle$ to show the observable, the residual observable is $\Delta\bar\Ob = \Ob-\sum_a\overline{w}_a(\Ob)\mu_a$, which is linear in $\Ob$. To keep the expansion compact, we also define the bilinear form $\G(A,B) \equiv \langle A,B\rangle+\Tr(A)\Tr(B)$ on Hermitian operators. The Haar-test error then becomes
\begin{equation}
    \beta^2_\Ob = \frac{\G(\Delta\bar\Ob,\Delta\bar\Ob)}{d(d+1)}
    = \frac{1}{d(d+1)}
    \Bigg[\G(\Ob,\Ob)-2\sum_a \overline{w}_a(\Ob)\G(\mu_a,\Ob) + \sum_{a,c}\overline{w}_a(\Ob)\overline{w}_c(\Ob)\G(\mu_a,\mu_c)\Bigg].
    \label{eq-app:bias-bilinear}
\end{equation}
For the Haar POVM prior of \cref{eq:haar-povm-prior,eq:haar-povm-prior-inv}, the weights are
\begin{equation*}
    \overline{w}_a(\Ob)=(\nout+1)\left[\langle\mu_a,\Ob\rangle-\frac{\Tr(\Ob)}{d+1}\Tr(\mu_a)\right].
\end{equation*}
We evaluate the traceless and identity contributions separately.

\subsubsection{Traceless component}

For the traceless component $\T$ of \cref{eq-app:center-obs}, $\Tr(\T)=0$, $\Tr(\T^2)=(d-1)/d$ and $\overline{\F}_\mu^{-1}(\T)=(\nout+1)\T$. Substituting into \cref{eq-app:bias-bilinear} gives
\begin{equation*}
    \beta^2_\T
    = \frac{d-1}{d^2(d+1)}
    - \frac{2(\nout+1)}{d(d+1)} \sum_a\langle\mu_a,\T\rangle^2
    + \frac{(\nout+1)^2}{d(d+1)}
    \sum_{a,c} \langle\mu_a,\T\rangle \langle\mu_c,\T\rangle \G(\mu_a,\mu_c).
\end{equation*}
Averaging over an independent Haar-random projector direction $\phi$ gives
\begin{equation*}
    \E_\phi[ \langle A,\T\rangle \langle B,\T\rangle ]
    = \frac{1}{d(d+1)}
    \left[ \langle A,B\rangle -\frac{\Tr(A)\Tr(B)}{d} \right].
\end{equation*}
Applying this identity to the measurement effects yields
\begin{align*}
    \E_\phi[\beta^2_\T]
    = \frac{d-1}{d^2(d+1)} \Bigg( 1
    &- \frac{2(\nout+1)}{d^2-1}
    \sum_a \left[ \langle\mu_a,\mu_a\rangle -\frac{\Tr(\mu_a)^2}{d} \right] \\
    &+ \frac{(\nout+1)^2}{d^2-1} \sum_{a,c}
    \left[ \langle\mu_a,\mu_c\rangle -\frac{\Tr(\mu_a)\Tr(\mu_c)}{d} \right]
    \left[ \langle\mu_a,\mu_c\rangle + \Tr(\mu_a)\Tr(\mu_c) \right]\Bigg).
\end{align*}
For the Haar POVM, $\mu_a=\ket{k_a}\!\bra{k_a}$, so that $\Tr(\mu_a)=\norm{\ket{k_a}}^2$ and $\langle\mu_a,\mu_c\rangle = \lvert\braket{k_a}{k_c}\rvert^2$. Evaluating the corresponding moments of the Haar POVM ensemble using Weingarten calculus~\cite{2006collins,2010collins} gives
\begin{equation}
    \E_{\mu,\phi}[\beta^2_\T]
    = \frac{d-1}{d^2(d+1)}
    \frac{\nout^2(d^2+3d+3)+\nout(2d^2+2d+1)+d^2-d}
    {\nout(\nout+2)(\nout+3)}.
    \label{eq-app:bias-T}
\end{equation}
Setting $\nout=\alpha_\mu d^2$ and expanding at fixed $d$ gives
\begin{equation*}
    \E_{\mu,\phi}[\beta^2_\T] =
    \frac{C_{\T}(d)}{\alpha_\mu} + O(\alpha_\mu^{-2}),
    \qquad
    C_{\T}(d) \equiv \frac{(d-1)(d^2+3d+3)}{d^4(d+1)}.
\end{equation*}

\subsubsection{Rank-one projector}

\begin{table}[t]
    \centering
    \begin{tabular}{lc@{\hskip 0.8em}cccc}
        \toprule
        Excess & $d$ & $s_\mathrm{th}$ & $s_\mathrm{fit}$ & $C_\mathrm{th}$ & $C_\mathrm{fit}$ \\
        \midrule
        $\Delta_\sp$ & 2 & $-1$ & $-0.955$ & $0.2500$  & $0.2654$  \\
        $\Delta_\sp$ & 4 & $-1$ & $-1.095$ & $0.09375$ & $0.1084$  \\
        $\Delta_\sp$ & 8 & $-1$ & $-1.078$ & $0.02734$ & $0.02945$ \\
        \midrule
        $\Delta_\mup$ & 2 & $-1$ & $-0.902$ & $0.3056$  & $0.2836$  \\
        $\Delta_\mup$ & 4 & $-1$ & $-1.017$ & $0.07375$ & $0.07287$ \\
        $\Delta_\mup$ & 8 & $-1$ & $-0.993$ & $0.01731$ & $0.01711$ \\
        \bottomrule
    \end{tabular}
    \caption{Summary of the prior-excess fits for the rank-one projector $\proj$. At each grid point, the fitted quantity $f(x)$ is the arithmetic mean of the paired per-seed excesses over 100 independent realizations. Shaded bands in \cref{fig:prior-excess} represent the mean $\pm$ one standard error of the mean. Fits are performed in logarithmic space over $x\geq10$, with $x=\alpha_\rho$ for the state-prior excess and $x=\alpha_\mu$ for the POVM-prior excess. The exponent $s_\mathrm{fit}$ is obtained from the free fit $f=10^q x^s$. The coefficient $C_\mathrm{fit}$ is instead obtained by fixing $s_\mathrm{th}=-1$, namely $C_\mathrm{fit}=10^{\langle\log_{10}f+\log_{10}x\rangle}$, where the brackets denote an average over the fitted grid points. It is compared with the theoretical coefficient $C_\mathrm{th}$ of the predicted law $f=C_\mathrm{th}/x$.}
    \label{tab:mse-excess}
\end{table}

The rank-one projector decomposes into its identity and traceless components as $\proj=I/d+\T$. Since the residual observable is linear in $\Ob$, the residual of the projector is the sum of the residuals of $I/d$ and of $\T$. Since $\E_\phi[\T]=0$, the mixed term in the quadratic error vanishes after averaging over the random projector direction. Consequently,
\begin{equation*}
    \E_{\mu,\phi}[\beta^2_\phi]
    = \E_\mu[\beta^2_{I/d}] + \E_{\mu,\phi}[\beta^2_\T].
\end{equation*}
For the identity contribution, $\overline{\F}_\mu^{-1}(I/d) = \frac{\nout+1}{d(d+1)}I$, so that $\overline{w}_a(I/d) = \frac{\nout+1}{d(d+1)} \Tr(\mu_a)$.

Evaluating the corresponding moments of the Haar POVM ensemble~\cite{2006collins,2010collins} gives
\begin{equation*}
    \E_\mu[\beta^2_{I/d}] = \frac{(\nout-d)(d\nout-d+3\nout+1)}
    {d^2\nout(d+1)^2(\nout+2)(\nout+3)}.
\end{equation*}
Combining this expression with \cref{eq-app:bias-T}, we obtain
\begin{equation}
    \E_{\mu,\phi}[\beta^2_\phi] =
    \frac{C_\phi(d)}{\alpha_\mu} + O(\alpha_\mu^{-2}),
    \qquad
    C_\phi(d) = C_{\T}(d) + \frac{d+3}{d^4(d+1)^2}
    = \frac{d^3+3d^2+2d-2}{d^3(d+1)^2}.
    \label{eq-app:povm-prior-excess-phi}
\end{equation}
The leading high-statistics prediction is therefore $\Delta_{\mup,\,\phi} \simeq C_\phi(d)/\alpha_\mu$.

\Cref{fig:prior-excess}(b) and \cref{tab:mse-excess} support the predicted inverse-$\alpha_\mu$ dependence for the rank-one projector. The fitted amplitudes agree with the asymptotic coefficients to within approximately $1.2\%$ for $d=4,8$, while the deviation is larger, approximately $7.2\%$, for $d=2$.

\section{OT as an instrumental-variable estimator}
\label{app:iv}

In this appendix we show in \cref{app:iv-eiv} that pseudo-inverse training is an errors-in-variables regression, and, in \cref{app:iv-2sls}, that OT coincides with the standard estimator used to correct it, two-stage least squares with instrumental variables. We first recall the two problems in classical notation and then map them onto QLR.

\subsection{Pseudo-inverse training as errors in variables}
\label{app:iv-eiv}

In classical linear regression the labels depend linearly on regressors $\bs x_i\in\RR^{\nout}$, $y_i=\bw^T\bs x_i$, and the least-squares estimator is unbiased in the underparametrized regime whenever the regressors are known exactly~\cite{2009hastie}. In an errors-in-variables model the regressors are instead observed with noise, $\es{\bs x}_i=\bs x_i+\bs\eta_i$, with $\E[\bs\eta_i]=0$, covariance $\bSigma_{\eta,i}$, and noise independent of the labels~\cite{1987fuller,2010buonaccorsi,2006carroll,2007soderstrom}. Collecting the observed variables as the columns of $\es{\bs X}\in\RR^{\nout\times\ntr}$, with $\ntr>\nout$ and $\es{\bs X}$ of full row rank, the least-squares estimator is
\begin{equation*}
    \hw = (\es{\bs X}^T)^+\by
    = \left(\frac{1}{\ntr}\es{\bs X}\es{\bs X}^T\right)^{-1}
      \left(\frac{1}{\ntr}\es{\bs X}\by\right).
\end{equation*}
As $\ntr\to\infty$, the difference between the cross terms $\es{\bs X}\by/\ntr$ and $\bs X\by/\ntr$ vanishes, because the noise has zero mean and is independent of the labels, while the Gram matrix acquires the average noise covariance $\overline{\bSigma}_\eta=\ntr^{-1}\sum_i\bSigma_{\eta,i}$,
\begin{equation}
    \hw \longrightarrow
    \left(\frac{1}{\ntr}\bs X\bs X^T+\overline{\bSigma}_\eta\right)^{-1}
    \frac{1}{\ntr}\bs X\by.
    \label{eq-app:iv-attenuation}
\end{equation}
The noise covariance acts as an implicit Tikhonov regularizer and biases the estimator towards zero. This is the attenuation bias, or regression dilution, of errors-in-variables models~\cite{1987fuller,2010buonaccorsi,2012loh}. It does not vanish with the size of the dataset and, within least squares, it is removed only by reducing the noise on the regressors.

Pseudo-inverse training has exactly this structure. The regressors are the outcome probabilities, $\bs x_i=\bp_i$, observed through the empirical frequencies $\es{\bs x}_i=\hbp_i$, and the labels $y_i=\langle\Ob,\rho_i\rangle$ are exact. By \cref{eq:multinomial-moments}, the regressor noise has zero mean and covariance $\bSigma_i/N$, so that \cref{eq-app:iv-attenuation} becomes
\begin{equation*}
    \hw_\pinv \longrightarrow
    \left(\frac{1}{\ntr}\bP\bP^T+\frac{\overline{\bSigma}}{N}\right)^{-1}
    \frac{1}{\ntr}\bP\by,
    \qquad \overline{\bSigma}=\frac{1}{\ntr}\sum_{i=1}^{\ntr}\bSigma_i ,
\end{equation*}
which differs from the ideal readout $\bw_\ex$ by a bias of order $1/N$, independent of $\ntr$. This bias is the origin of the $b^2/N^2$ floor in \cref{eq:mse-pinv}. QLR differs from the classical errors-in-variables setting in two respects~\cite{2026lomonaco}. First, the noise strength $1/N$ is set by the number of shots per state, so it is chosen by the experimenter rather than fixed by the data. Second, the exact Gram matrix $\bP\bP^T$ has rank at most $d^2<\nout$, so the shot noise also populates directions in which the exact features have no support. For this reason the exact result of Ref.~\cite{2026lomonaco} involves the Schur complement of $\overline{\bSigma}$ on the support of $\bP$ rather than $\overline{\bSigma}$ itself, but the mechanism is the same.

\subsection{Instrumental variables and two-stage least squares}
\label{app:iv-2sls}

The classical remedy for errors in variables is the method of instrumental variables~\cite{1954durbin,2001hausman,2001angrist,2009angrist,2010wooldridge}. An instrument $\bs z_i\in\RR^{k}$ is an additional observed quantity that is correlated with the true regressors but independent of their noise, and of the label noise if present. In the linear case the regressors are related to the instruments by a first-stage model $\bs x_i=\bs\Pi\bs z_i+\bs u_i$. Collecting the instruments as the columns of $\bs Z\in\RR^{k\times\ntr}$, the two-stage least-squares (2SLS) estimator~\cite{1953theil,1957basmann,2010wooldridge} proceeds as follows:
\begin{enumerate}
    \item regress the noisy regressors on the instruments, $\es{\bs\Pi}=\es{\bs X}\bs Z^+$, and form the fitted regressors $\bs X_\mathrm{fit}=\es{\bs\Pi}\bs Z$;
    \item regress the labels on the fitted regressors, $\hw_{\mathrm{2SLS}}=(\bs X_\mathrm{fit}^T)^+\by$.
\end{enumerate}
The fitted regressors are the projections of $\es{\bs X}$ onto the row space of $\bs Z$. Since the instruments are independent of the noise, the projected noise vanishes as $\ntr\to\infty$, and 2SLS is consistent where ordinary least squares is not~\cite{2010wooldridge,2009angrist}. Its residual bias is a finite-sample effect that decreases with $\ntr$, and it is small when the instruments are strongly correlated with the regressors~\cite{1995bound,1997staiger}.

In QLR the known training states are instruments for the noisy frequencies. We take $\bs z_i=\bv_{\rho_i}$, so that $\bs Z=\Mrho$ and $k=d^2$. By the Born rule the first-stage model holds exactly, $\bp_i=\Mmu^T\bv_{\rho_i}$, with $\bs\Pi=\Mmu^T$ and $\bs u_i=0$, and the instruments are independent of the shot noise because they are prepared by the experimenter. The two stages of 2SLS then read:
\begin{enumerate}
    \item the first stage is the measurement estimate of OT, \cref{eq:explicit_estimator_OT}, $\es{\bs\Pi}=\PN\Mrho^+=\hatMmu^T$, with fitted regressors $\bs X_\mathrm{fit}=\hatMmu^T\Mrho$;
    \item the second stage gives
    \begin{equation}
        \hw_{\mathrm{2SLS}}=\left(\Mrho^T\hatMmu\right)^+\by
        =\left(\Mrho^T\hatMmu\right)^+\Mrho^T\bv_\Ob .
        \label{eq-app:iv-2sls-qlr}
    \end{equation}
\end{enumerate}
If the training states span the operator space, $\Mrho$ has full row rank $d^2$, and if the estimated effects are IC, $\hatMmu$ has full row rank $d^2$. For such a full-rank factorization the reverse-order law holds, $(\Mrho^T\hatMmu)^+=\hatMmu^+(\Mrho^T)^+$~\cite{2012barata}, and $(\Mrho^T)^+\Mrho^T=\Mrho\Mrho^+=I$. \Cref{eq-app:iv-2sls-qlr} then reduces to
\begin{equation*}
    \hw_{\mathrm{2SLS}}=\hatMmu^+(\Mrho^T)^+\Mrho^T\bv_\Ob=\hatMmu^+\bv_\Ob=\hw_\OT,
\end{equation*}
the OT readout of \cref{eq:explicit_estimator_OT}. OT is therefore the 2SLS estimator of the QLR problem, with the training states as instruments.

The quantum setting is a particularly favorable instance of instrumental-variable regression. The instruments are known exactly and chosen by the experimenter, the first stage is exactly linear, and its coefficients $\bs\Pi=\Mmu^T$ are the effective measurement itself, so that the first stage is detector tomography with the training states as probes~\cite{2001fiurasek,2009lundeen,2017motka}. The strength of the instruments is set by the conditioning of the state frame $\Mrho\Mrho^T$, which is good for spanning ensembles such as Haar-random states. Finally, since the target observable is known, the second stage does not require the labels and can be written directly in terms of $\bv_\Ob$. A similar distinction appears in quantum state tomography with uncalibrated detectors: standard tomography first reconstructs the detector and then uses it, while data-pattern tomography fits the data directly with the measured patterns~\cite{2014motka,2017motka}.

\end{document}